\documentclass[8pt,letterpaper]{article}

\usepackage{graphics}

\usepackage{graphicx}
\usepackage{epsfig}
\usepackage{caption} 
\usepackage{subcaption} 

\usepackage{float} 
\usepackage{indentfirst}

\usepackage{amssymb}
\usepackage{amsthm}

\usepackage{amsmath}
\usepackage{lineno}
\usepackage{authblk}
\usepackage{subfig}
\usepackage{booktabs}

\begin{document}

\title{\Large Foam-like compression behavior of fibrin networks}

\author[1,+]{O.V. Kim}
\author[2,+]{Xiaojun Liang}
\author[3]{Rustem I. Litvinov}
\author[3]{John W. Weisel}
\author[1,4]{Mark S. Alber}
\author[2]{Prashant K. Purohit\thanks{purohit@seas.upenn.edu}}

\affil[+]{These two authors contributed equally}
\affil[1]{Department of Applied and Computational Mathematics and Statistics, University of Notre Dame, Notre Dame, Indiana}
\affil[2]{Department of Mechanical Engineering and Applied Mechanics, University of Pennsylvania, Philadelphia, PA}
\affil[3]{Department of Cell and Developmental Biology, Perelman School of Medicine, University of Pennsylvania, Philadelphia, Pennsylvania}
\affil[4]{Department of Medicine, Indiana University School of Medicine, Indianapolis}

\date{}
\maketitle

\begin{abstract}
{The rheological properties of fibrin networks have been of long-standing interest. As such there is a wealth of studies of their shear and tensile responses, but their compressive behavior remains unexplored. Here, by characterization of the network structure with synchronous measurement of the fibrin storage and loss moduli at increasing degrees of compression, we show that the compressive behavior of fibrin networks is similar to that of cellular solids. A non-linear stress-strain response of fibrin consists of three regimes: 1) an initial linear regime, in which most fibers are straight, 2) a plateau regime, in which more and more fibers buckle and collapse, and 3) a markedly non-linear regime, in which network densification occurs {{by bending of buckled fibers}} and inter-fiber contacts. Importantly, the spatially non-uniform network deformation included formation of a moving ``compression front" along the axis of strain, which segregated the fibrin network into compartments with different fiber densities and structure. The Young's modulus of the linear phase depends quadratically on the fibrin volume fraction while that in the densified phase depends cubically on it. The viscoelastic plateau regime corresponds to a mixture of these two phases in which the fractions of the two phases change during compression. We model this regime using a continuum theory of phase transitions and analytically predict the storage and loss moduli which are in good agreement with the experimental data. Our work shows that fibrin networks are a member of a broad class of natural cellular materials which includes cancellous bone, wood and cork.
}
\end{abstract}

\section{Introduction}
Fibrin is a polymer made from the blood plasma precursor protein fibrinogen. Fibrin forms a porous network of branching fibers that provides the scaffold of protective hemostatic blood clots and pathological obstructive thrombi. The structure and properties of the fibrin, including mechanical resilience to forces of blood flow and clot contraction, determine the course of pathological conditions, such as coronary heart disease, ischemic stroke, or bleeding \cite{Bio,b}. Hence, the tensile and shear behaviors of fibrin networks have been studied for a long time \cite{FerryBook, Bio}. Recent experimental studies have also focused on the tensile properties of a single fibrin fiber including the molecular unfolding of fibrin molecules at large deformations \cite{Falvo1,Falvo2,Falvo3,Falvo4,Falvo5}. The tensile behavior of single fibrin fibers can be used as an input to network models to predict the tensile behavior of the networks. We did this recently using the eight-chain model and were able to predict the mechanics of fibrin networks under tension \cite{Science,Acta,Mac1,Mac2,Mac3}. The shear behavior of fibrin networks has also been studied both experimentally and theoretically \cite{Janmey, Mac1, Mac2, Mac3, weiselbiophys}. For networks with thin fibrin fibers the shear behavior can also be explained in terms of the entropic elasticity of single fibrin fibers \cite{Mac1, Mac3}. 

\begin{figure}[]
\centering
\includegraphics[scale=0.20]{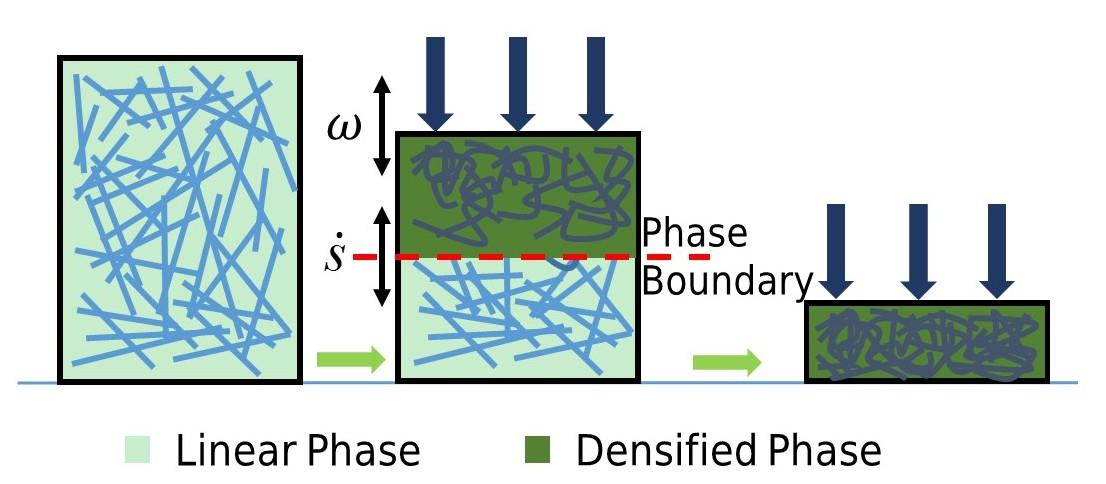} 
\includegraphics[scale=0.25]{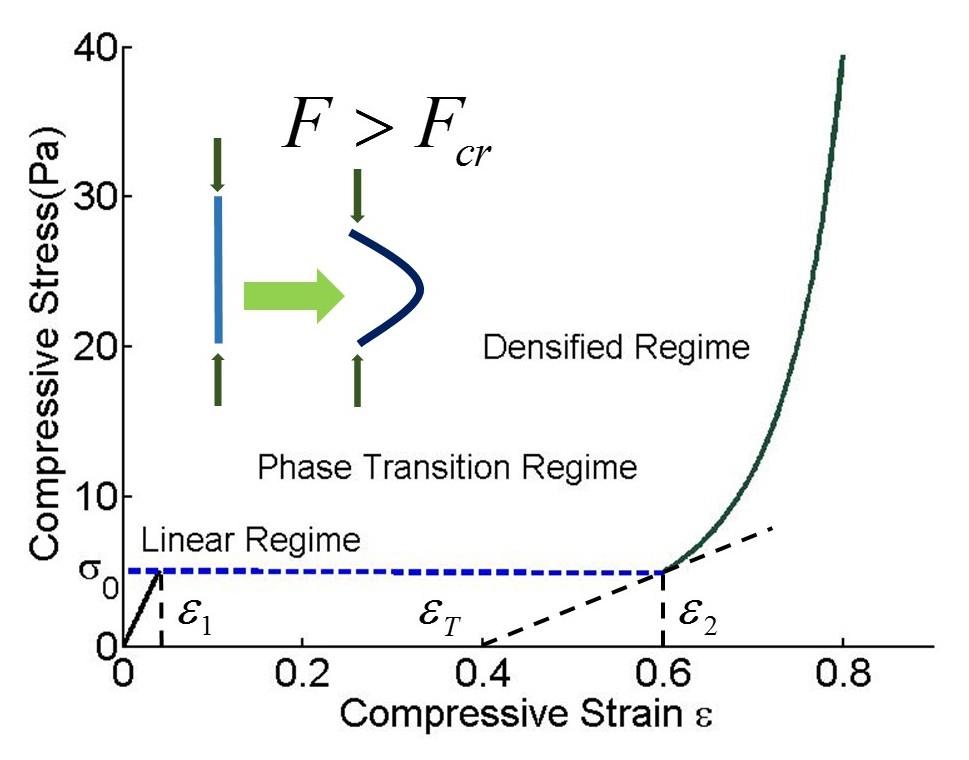} 
\caption{A fiber network under compression. (a) The top panel is a cartoon, schematically showing the high strain phase with a densified network on the right, the low strain phase with straight fibers on the left and a mixture of these two phases in the middle. Here $\omega$ indicates the frequency of applied oscillation and $\dot s$ is the rate of change of phase fraction due to this oscillation. In the middle figure a sharp phase boundary separates the linear phase in the bottom from the densified phase on the top. The fraction of densified phase increases with increasing compressive strain. (b) The bottom panel quantitatively shows the tri-phasic stress-strain response of such a material assuming quasi-static loading. Theoretical curves are plotted according to Eqns. (\ref{E1}) and (\ref{E2}). We have used $l=1.32\mu \text{m}$, $d=0.22\mu \text{m}$, $\nu = 0.1\mu \text{m}^{-3}$, and the two coefficients $n = 3$, $k =0.5$ corresponding to a fibrin network. }
\label{pha}
\end{figure}

Even though tension and shear behaviors of fibrin networks have been studied in detail, compression of these networks remains largely unexplored. Recently, Kim {\it et. al.}~\cite{Oleg} found that the non-linear stress-strain curves of fibrin networks in compression have three distinct regimes, similar to the curves observed in networks of carbon nanotubes \cite{Greer}. There is an initial linear response which is expected of any elastic solid for small strains, followed by a stress plateau that begins when fibers in the network start to buckle and continues until a majority of them have done so. Then, network densification occurs creating a large number of contacts between the fibers as the stress increases steeply. This tri-phasic mechanical behavior could presumably be modeled using the theory of foams and other cellular solids \cite{foam} as shown schematically in Figure~\ref{pha}. The initial shear and compression moduli of foams in the linear regime can be computed in terms of the Young's modulus of the foam material and the geometric parameters of the foam \cite{foam}. These moduli can also be connected to the moduli of fiber networks as has been done by \cite{PRL,PRL2,PRE}. But, for large compressions these moduli are no longer sufficient and we must use more sophisticated theories. 

Our goal in this paper is to further study the mechanical and structural responses of fibrin to compression by a combination of experiment and theory. In particular, from precise dynamic rheological and microscopic measurements we have developed a model for the compression behavior of fibrin networks based on the theory of foams and other cellular solids. First, the compression stress-strain curve was measured in experiment along with the loss and storage moduli at various strains with simultaneous 3D visualization of the fibrin network during the deformation. In addition to the non-linear behavior of storage and loss moduli, we were able to reveal the non-uniformity of the compressive deformation with formation of a ``compression front'' or ``phase boundary'' along the axis of compressive strain (see Figure~\ref{pha}) \cite{PB1,PB2}. Next, we described how the non-linear rheological behavior could be explained by viewing the fibrin network as a ``cellular'' solid that could exist in two phases -- the low-strain phase in which the fibers are mostly straight, and a high-strain phase in which the fibers are mostly buckled. In displacement-controlled experiments such a solid has a stress plateau  when there is a mixture of both phases. The fraction of each phase at equilibrium  is determined by a lever rule~\cite{Landau}. However, the storage and loss moduli at various degrees of compression in our fibrin networks were obtained in the experiments by performing small oscillations around a certain strain. These oscillations changed the fractions of the two phases but they were not necessarily quasi-static. {We treated} these two phases using a continuum theory of phase transitions in which a kinetic equation relates the rate of change of the fraction of the phases to the stress \cite{BookPhase}. In this way we explained the storage and loss moduli of the networks while accounting for the contributions of the moving phase boundary as well as the viscoelasticity of each phase.

In this paper we first provide the details of the experimental procedures and present a description of the obtained data on controlled fibrin compression followed by a comprehensive theoretical analysis. After a short summary of the well-known results for cellular solids we show how this approach can be applied to fiber networks. We also describe the constitutive laws for fiber networks under large compression that have been studied using experiments and theory for several decades \cite{Wyk, Toll}. We finally show how these results can be generalized and combined with a continuum theory for phase transitions to obtain the storage and loss moduli of networks under compression. {{The novelties of this paper are (a) experimental visualization of a propagating phase boundary in compressed fibrin networks, (b) an explanation for the trends in storage and loss moduli of the compressed networks in terms of a theory of phase transitions, and (c) first application of foam theory to a bio-polymer network. Our analysis leads to some new experimentally testable predictions about the 
dependence of the mechanical properties of the clot on network parameters that have been summarized in the discussion section.}}

\section{EXPERIMENTAL PROCEDURES}

\subsection{Fibrin Clot Formation}
Fibrin clots were formed by adding human $\alpha$-thrombin (1 U/mL final concentration) to pooled human citrated platelet-poor plasma in the presence of CaCl$_2$ (40 mM final concentration). Before clotting, Alexa-Fluor 488-labeled human fibrinogen (Molecular Probes, Grand Island, NY) was added to the plasma samples ($5\%$ of total volume, 0.08 mg/mL final concentration) to visualize fibrin network structure in a fluorescence confocal microscope. The clots were allowed to form at room temperature for 90-100 minutes to ensure covalent cross-linking by factor XIIIa. Ten fibrin clots from different plasma samples were prepared as described and used for compression studies with combined structural and rheological examination. 

\subsection{Visualization of Fibrin Networks and Image Analysis}
The fluorescent confocal microscopy and strain-controlled oscillatory measurements were performed synchronously using a home-built ``confocal rheoscope'' which combined Bohlin Gemini rheometer with Nikon Eclipse 200/Confocal VT Eye microscope \cite{Oleg, Wen}. The plasma clots were allowed to form between horizontal rheometer plates separated by a distance of 650-750 $\mu$m. The sample volume was 300-375 $\mu$l. The upper moving acrylic plate had a 20-mm diameter and the base motionless plate was made of a 22-mm-diameter round transparent microscope glass cover slip to perform confocal imaging of the network during rheological measurements.
 
To improve the image quality, in a separate series of experiments uncompressed and compressed fibrin clots were analyzed using the Zeiss LSM510 META NLO laser scanning confocal microscope to get higher resolution images with z-stacks spanning the entire clot thickness. We used Plan Apo 40x water immersion objective lens (NA 1.2).  A laser beam with wave length of 488 nm was used for fluorescent imaging. The distance between slices in each z-stack was 0.5 $\mu$m. The image resolution of each slice was 1024x1024 pixels. 

The samples were subsequently processed and analyzed quantitatively using the automated program as described in \cite{nex1}. Briefly, a new two-stage method was used to extract the structure of the network in the 3D confocal microscopy images and to characterize the structure by vertices (end points and branch points of the network) and edges (fiber network segments) in a 3D graph. A detailed description of the algorithm can be found elsewhere \cite{nex1}.

\subsection{Controlled Compression of Fibrin Clots}
For confocal microscopy there were two types of fibrin clots compressed using a computerized AR-G2 Rheometer (TA Instruments, New Castle, DE). First, the 650-750-$\mu$m-thick fresh hydrated fibrin clots were formed directly between the rheometer plates and compressed vertically down to 1/10 of their initial thickness. 12 clots from various plasma samples were tested. The stepwise compression of the clots was performed in 25-$\mu$m and 50-$\mu$m steps at the rate of 30 $\mu$m/s, as the downward force was applied by the upper rheometer plate on the top surface of the clot. In the other type of compression experiments the 150-$\mu$m-thick fibrin clots were formed in a flow chamber built of a microscope glass slide (bottom) and a glass cover slip (top) separated by an adhesive and highly plastic putty, which permitted irreversible compression of the clot samples within the chamber placed between the rheometer plates. In both types of experiments the fully polymerized and cross-linked plasma clots were compressed to a different degree followed by confocal microscopy. {{We typically
waited 2 minutes after each step before performing confocal microscopy.} A compressive strain (or degree of compression), $\varepsilon$ was defined as the absolute fractional decrease in fibrin clot thickness $\varepsilon = \left| {\Delta L/{L_0}} \right|$, where $\Delta L=L-L_0$, and $L_0$ and $L$ are the initial and reduced thickness dimensions of the uncompressed and compressed clots, respectively.}

\subsection{Oscillatory Shear Rheometry of Compressed Fibrin Clots}
Changes of viscoelastic properties of fibrin clots during compression were measured by oscillating rheometry using a Bohlin Gemini rheometer (Malvern Instruments, Westborough, MA) with parallel plate geometry. Measurements were performed with a fixed oscillating 0.5$\%$ strain amplitude at a frequency of 1 Hz to produce a linear stress-strain response to imposed shear.  The oscillations were initiated immediately after each subsequent step of compression and lasted for 300 s. In each test, the storage ($G'$) and loss ($G''$) moduli were measured in combination with confocal microscopy of the fibrin clot structure. The elastic response of the deformed clot to applied shear was characterized by the storage modulus, $G'$, which represents the stored energy and is defined as $G'=
\left(\tau_0/\gamma_0\right)\cos(\delta)$. Here, $\gamma_0$ is a (small) strain amplitude, $\tau_0$ is a shear stress amplitude, and $\delta$ is a phase shift in the shear stress with respect to applied oscillatory strain. The viscous response of the deformed clot to applied shear was measured by the shear loss modulus, $G''$, corresponding to the energy dissipated as heat, and calculated as $G''= \left(\tau_0/\gamma_0\right)\sin(\delta)$.

\begin{figure*}[h]
\centering
\includegraphics[width=5in]{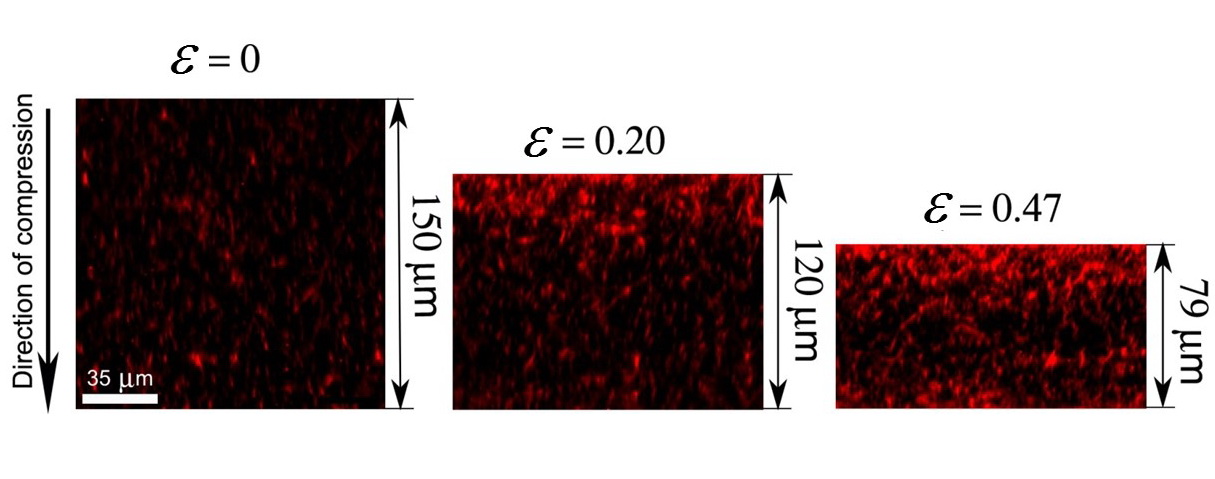}
\caption{The ``compression front'' or ``phase boundary'' is formed in response to vertical deformation of a fibrin clot. {{To provide visual examination of the changing clot structure, the distribution of fluorescence intensity in the XZ plane is shown in the same fibrin clot, uncompressed (left), about 20$\%$ compressed (center), and about 50$\%$ compressed (right).}} As the degree of compression increases, the fluorescence intensity (reflecting the network density) displays a gradient along the direction of compression. Here the compressive strain $\varepsilon$ is a degree of compression defined as $\varepsilon = \left| {\Delta L/{L_0}} \right|$, where $\Delta L=L-L_0$, and $L_0$ and $L$ are the initial and reduced thickness dimensions of the uncompressed and compressed clots, respectively. {{The dashed lines outline the top (I), middle (II) and bottom (III) layers of the compressed clot, respectively. A horizontal arrow indicates the position of the compression front. }} }
\label{3d}
\end{figure*}

\section{EXPERIMENTAL RESULTS}
\subsection{Compression-Induced Non-Uniformity of the Fibrin Network Structure }
\begin{figure}[]
\centering
\includegraphics[width=3in]{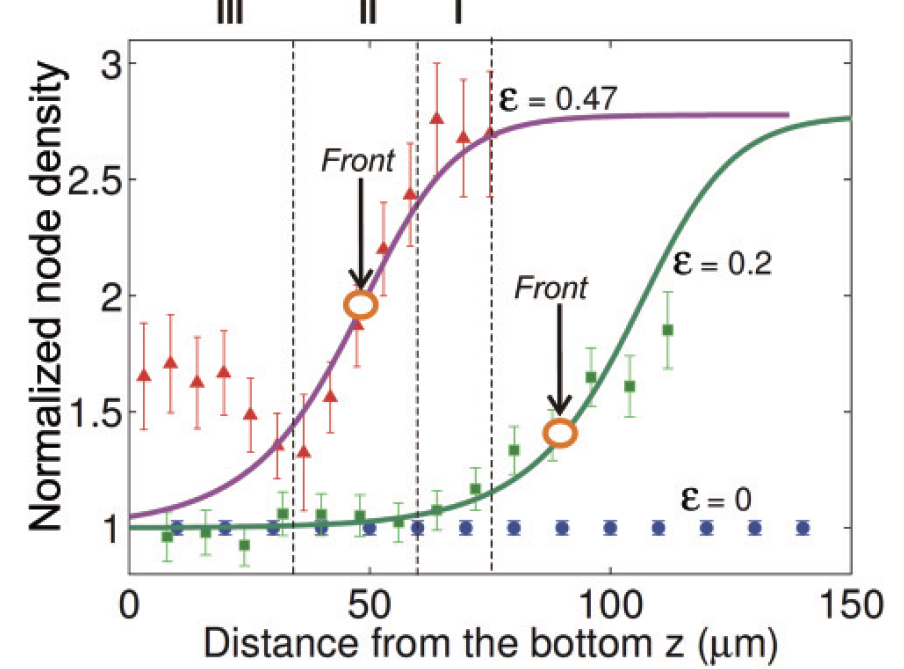}
\includegraphics[width=3in]{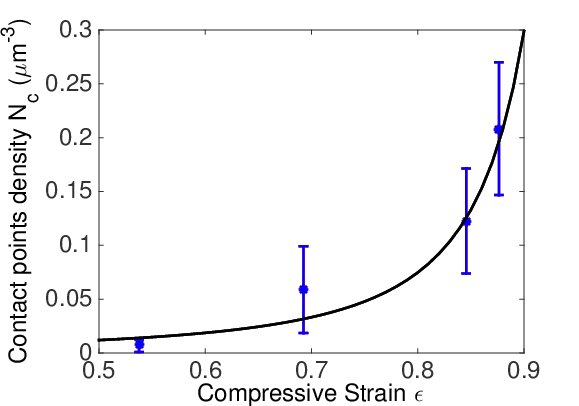}
\caption{(a)The node density of the fibrin network as a function of height from the bottom of the sample based on 3D reconstruction of the network, {{uncompressed ($\varepsilon = 0$) and compressed to different degrees ($\varepsilon = 0.2$ and $\varepsilon = 0.47$). The node density was normalized with respect to the node density of the uncompressed clot. Note that the node density increases from bottom to top. The curve is flatter near the top and bottom of the sample and has a larger gradient near the middle.  This suggests that the fibers are buckled near the top and straight near the bottom. The symbols represent experimental data (MSD, n=4) and we have modeled the regime of high gradient as a compression front or phase boundary. The lines represent fitting curves made using Eqns. (\ref{eq:frontfit}) and (\ref{eq:defheight}) together with the experimental data with {{fitting parameters}} $a=b=0.2$, $c=30$, and $Z_0 = 42$ for red curve, $Z_0 = 100$ for green curve respectively.}} {{$Z_{0}$ represents the center of the phase boundary in the 
reference configuration. The decrease of $Z_{0}$ shows that}} the front moves downwards in response to increased compression. {{The top (I), middle (II) and bottom (III) layers of the compressed clot ($\varepsilon=0.47$) are separated by the vertical dashed lines that correspond to the layers shown in Fig. 2. Large circles with vertical arrows indicate the position of the compression front.}} (b) Experimental data of node density under large compression is fitted by Eqn. (\ref{Nc}). This justifies our use of Eqn. (\ref{E2}) for the densified network. We have used $l=1.32\mu \text{m}$, $d=0.22\mu \text{m}$, $\nu = 0.1\mu \text{m}^{-3}$.}
\label{phac}
\end{figure}

Visual examination of the fluorescent confocal images of the fibrin clots during
compression revealed marked non-uniformity of the network structure. Here we define the XYZ-coordinate system as Z-axis along the vertical direction 
that is opposed to the applied compression. The XZ-plane sections of the fibrin networks clearly showed the appearance and progression of a gradient in the fluorescence intensity in the direction of compression with a more or less sharp boundary between the areas with different intensities (Figure \ref{3d}).  To precisely quantify the non-uniformity of the network density, we computationally reconstructed and analyzed the 3D images of a 150-$\mu$m-thick uncompressed fibrin clot and the same clot after 20$\%$ and 50$\%$ compression by segregating each image into 14 equal horizontal {sublayers} spanning the entire height of the network. The thickness of the {sub}layers at various degrees of compression was different depending on the height of the clot.  We quantified the uniformity of network densification as the node density in each layer and plotted it against the distance from the bottom of the clot compressed from above (Figure \ref{phac}). Our results showed that during vertical compression of a fibrin network a much higher degree of node densification was observed closer to the top of a clot, while the node and network density decreased towards the bottom. 
The top, middle and bottom layers of the compressed clots are indicated in Figure \ref{3d} and Figure \ref{phac}. As the degree of compression increased from $\varepsilon=0$ to $\varepsilon=0.2$ and $\varepsilon=0.47$, the node density of the upper network layers increased by a factor of 1.8 and 2.3, respectively. At the same degrees of compression ($\varepsilon=0.2$ and $\varepsilon=0.47$) the node density of the middle layers of the network increased by only 1.1 and 1.2, respectively. The density of the bottom portion of the network did not change at $\varepsilon=0.2$, but increased by 1.2 for $\varepsilon = 0.47$ {{as is apparent from the diffuse region near the bottom of the clot in Figure~\ref{3d} (right panel). However, this does
not contradict the presence of a moving compression front from the top of the clot. Since the width of the front given by the parameter $c$ in 
Eqn. (\ref{eq:frontfit}) is about $30\mu$m 
and its center is located at $Z_{0} = 42\mu$m, the bottom portions of the network likely experience a non-zero strain that pushes them against the glass 
surface causing a moderate increase in fluorescence intensity.} }

\begin{figure}[h]
\centering
\includegraphics[width=3in]{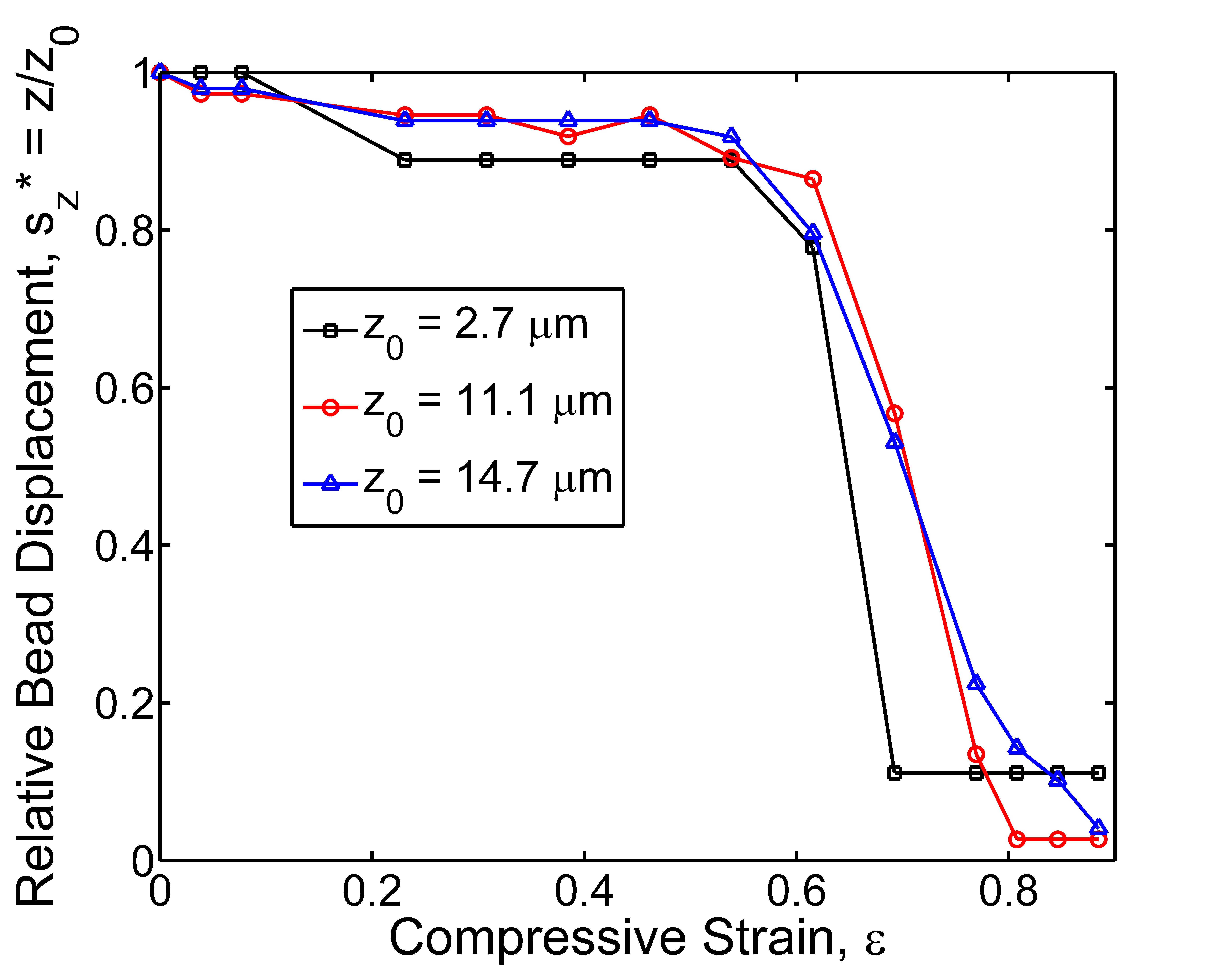}
\caption{{{The absolute vertical distance from the bead to the bottom of the clot along the direction of network compression for various locations in the same clot compressed from ${ {\varepsilon} } = 0-0.9$.}} The network is characterized by $z$ - the bead distance from the bottom at a certain degree of compression and $z_0$ - the bead initial distance from the bottom of the network. $2\mu m$ polystyrene fluorescent beads were tracked during the compressive deformation in a clot volume of $35.8 \times 35.8 \times 25.5 \mu m$ in the bottom portion of the clot. Densification occurs at the bottom of the clot only at high compressive strain, likely coinciding with the arrival of the compression front at the bottom. Before the arrival of the compression front the density in the bottom of the clot remains almost independent of strain. This is what we expect in the plateau regime when the network consists of a mixture of two phases that accommodate the increasing compressive strain simply by advancing the compression front.}
\label{rd}
\end{figure}

To confirm the structural non-uniformities in fibrin clots produced by compressive loads, we also complemented the 3D structural analysis of compressed fibrin networks with an experiment to follow the displacement of microscopic fluorescent beads embedded into the network in response to clot compression. 2-$\mu$m polystyrene fluorescent beads were tracked during the compressive deformation in a clot volume of $35.8 \times 35.8 \times 25.5 \mu$m in the bottom portion of the clot. {{The compression was applied in $25\mu$m steps. After each step we waited 2 minutes before measuring the displacements of
the beads.}} We found that the beads' position changed non-linearly in the Z-direction as a function of compressive strain (Figure \ref{rd}), indicating the non-uniform compression of the network. In the case of spatially uniform compression, one would expect to observe a linear displacement of beads with compression irrespective of their initial position and the distance from the bottom of a clot. However, in our experiments the beads exhibited less than 20$\%$ relative displacements when the clot was exposed to compressive strains from $\varepsilon = 0$ to $\varepsilon = 0.6$, but revealed a rapid descent toward the bottom of the clot at $\varepsilon > 0.6$. {{To be more specific, when the compressive strain reached $0.6$, the absolute distances changed by 
$2, 1.5$ and $0.6\mu$m for the beads initially located at $14.7, 11.1$ and $2.7\mu$m over the bottom, respectively. At a compressive strain of $0.7$ the 
bead, which was the closest to the bottom of the clot, stopped moving because it had reached the surface. As the strain increased, other beads moved 
downwards until they also reached the bottom of the clot. The trajectory of all the beads turned downward at a similar compressive strain because they are all
located within $15\mu$m of the bottom while compression is applied in $25\mu$m steps and the width of the front is about $30\mu$m (see Figure \ref{3d} and Figure \ref{phac}).} }The nonlinear response of bead displacement to network compression indicated that the deformation of the network occurred non-uniformly with the top layers being compressed earlier and stronger than the lower network portions. As a result, the beads did 
not undergo large displacements at low strains but their vertical position started to change drastically at about $\varepsilon > 0.6$ as the boundary 
between the relatively compact (upper) and loose (lower) portions of the network approached the bottom of the clot.

Thus, both the structural analysis and bead tracking provide evidence for the formation and propagation of a ``compression front" inside the network as more dense upper layers overtake the less dense bottom portions of the fibrin clot during progressive compression. Such a front can be represented by a strain
profile given as:
\begin{equation} \label{eq:frontfit}
 \varepsilon = a + b\tanh(\frac{Z - Z_{0}}{c}),
\end{equation}
\begin{equation} \label{eq:defheight}
 z(Z) = d + (1- a)Z - {b}{c}\log\left(\cosh(\frac{Z - Z_{0}}{c})\right),
\end{equation}
where $Z$ is the original height of a material point in the unstressed (reference) configuration, $Z_{0}$ is the location of the center of the front in the 
unstressed configuration, and $z$ is the deformed height that is measured in the experiment. We got Eqn. (\ref{eq:defheight}) by integrating 
${dz}/{dZ} = 1- \varepsilon$ with boundary condition $z(0) = 0$, as in our experiments. {{Capital $Z_{0}$ cannot be measured in our experiment but
using Eqn. (\ref{eq:defheight}) the change in $Z_{0}$ can be obtained from the experimentally measured changes in the location of the phase boundary as a 
function of $z$ as explained below.}}
The network density is proportional to $1/(1 - \varepsilon)$ as discussed in section \ref{short}. 
Thus, we can plot the network or node density as a function of $z$. Since the minimum and maximum values of the strain in 
Eqn. (\ref{eq:frontfit}) are $a \pm b$, we expect that $2b$ should approximately equal the transformation strain or the difference in strains between the 
high-strain and low-strain phases at the plateau stress $\sigma_{0}$ (see Figure \ref{pha}). $a$ should be the strain at the center of the front, $c$ should be a measure of the 
width of the front, and $d$ is obtained by enforcing $z(0) = 0$ (see analysis in Appendix). Keeping this in mind we chose reasonable values of 
$a$, $b$ and $c$ to fit the experimental data. The data and the fits are shown in Figure~\ref{phac}. We found that $2b \approx 0.4$, as 
expected to be the transformation strain shown in Figure \ref{pha}. Furthermore, $Z_{0}$ comes closer 
to the bottom of the sample when higher compression is applied, confirming that the ``compression front" moves downward in response to the loading. 

\subsection{Non-linear Mechanical and Structual Response of Fibrin Networks to Compression}
Structural alterations of fibrin network during compression underlie changes of 
shear viscoelasticity studied with rotational rheometry. The rheological measurements of fibrin clots with various initial shear storage and loss moduli were performed with two different compression steps (10 $\mu$m and 25 $\mu$m ) and at different degrees of compression. The experimental results are shown in Figures \ref{ws} and \ref{G} in combination with the theoretical curves. The normal stress, calculated as the force distributed over the area of the smaller rheometer plate was measured for each degree of compression, and presented as the plots of normal stress, $\sigma$, versus compressive strain, $\varepsilon$ (Figure \ref{ws}). The curves were almost linear at lower degrees of compression with a steep increase of the normal stress at higher degrees of compression ($\varepsilon>0.5$). The evolution of the shear viscoelastic properties of fibrin networks in response to compressive deformation displayed a typical tri-phasic behavior (Figure \ref{G}). In the beginning of compression ($\varepsilon <0.05 $), the shear elastic modulus of a fibrin network was almost constant followed by a long relative plateau up to about $\varepsilon =0.7-0.8$ compressive strain and finally dramatic increase at the highest compressive strains. The loss modulus followed the same trend (Figure \ref{G}). If the current rheological measurements are correlated with the structural analysis of the network dynamics performed earlier in \cite{Oleg}, the results could be described as follows. First, a linear viscoelastic response to compression was observed, in which most fibers are straight. Then, a plateau regime follows, accompanied by an increasing number of buckled fibers while the stress remains nearly constant as strain increases. Finally, we see a regime in which network densification occurs with a stress-strain response that is markedly non-linear and dominated by {{bending of fibers after buckling and inter-fiber contact. In the next section
we will quantify these observations using a model for foams. We emphasize that the experimental results reported here are substantially distinct from those 
reported earlier~\cite{Oleg} because in addition to the confirmed non-linear response, we were able to reveal the non-uniformity of the compressive deformation with 
formation of a ``compression front" or ``phase boundary" along the axis of compressive strain (see Figure~\ref{phac})~\cite{PB1, PB2}. }}
 
\section{THEORETICAL ANALYSIS OF THE EXPERIMENTAL DATA}
\subsection{Short Review of Cellular Solids} 
\label{short}
A foam is a porous low density structure consisting of cells or fibers made of a viscoelastic material. In our case these fibers are made of fibrin. If $E_s$ is Young's modulus of a single fibrin fiber, then the Young's modulus and shear modulus of the network or foam consisting of these fibers is given by \cite{foam}:
\begin{equation}
E_1 = E_s\phi^2_{{0}},
\label{E1}
\end{equation}
\begin{equation}
G_1 = \dfrac{3}{8}E_s\phi^2_{{0}},
\label{G1}
\end{equation}
where the important non-dimensional quantity $\phi = \pi \nu d^2 l/4$ is the volume fraction of fibrin in the network, $\nu$, $d$, and $l$ are respectively the number of fiber segments per unit volume, fiber diameter and fiber segment length {{between branch points. $\phi_{0}$ is the initial fiber
volume fraction before compression.}} These expressions are derived under the assumption of small strains so that the stress-strain relation of the foam can be approximated as linear. We will designate this phase of our fibrin network as the linear phase or the 
low-strain phase.

When this network is loaded in compression fibers will {{buckle \cite{foam}} when a critical force is reached}. From Euler's formula, the critical axial load that causes fiber buckling is
\begin{equation}
F_{cr} = \frac{n^2\pi^2E_sI}{l^2},
\label{Fcr}
\end{equation}
where $E_s,I$ and $l$ are the Young's modulus, area moment of inertia, and the length of the fiber {{between branch points}}, respectively. The coefficient $n$ is close to one and is determined by boundary conditions at the fiber ends. According to the analysis given in \cite{foam}, $n$ is taken to be $1$ because the fiber lengh $l$ varies over a range in random
networks. We can use this expression for the critical load to estimate the stress $\sigma_{0}$ at which buckling of fibers begins in the network. According to \cite{foam}, since, ${\sigma _0} \propto {F_{cr}}/{l^2} \propto {E_s}I/{l^4}$, we could write it as
\begin{equation}
{\sigma _0} = c{E_s}{\phi_{0}^2},
\end{equation}
where the coefficient $c$ is assumed to be 0.04 in \cite{foam}. The strain corresponding to the critical stress in the low-strain phase is easily computed as $\varepsilon_{1} = \sigma_{0}/E_1$. Another way to estimate $ \varepsilon_{1} $ is to use Eqn. (\ref{Fcr}) --  $\varepsilon_{1} \approx
 F_{cr}/(AE_s)$ where $A$ is fiber cross-sectional area. Both these methods give an estimate of $\varepsilon_1$ around $0.05$ and $\sigma_0$ around 5 $Pa$ 
as shown in Figure~\ref{pha} for the parameters in the experiments shown in Figures~\ref{ws} and \ref{G}. {{These ideas for estimating 
$\sigma_{0}$ can only be justified if the fibers in the network are not undulating due to thermal motion. The extent of thermal undulations depends on the
persistence length of the fibers and their length between branch-points. The persistence length of fibrin varies over a wide range depending on the 
conditions (thrombin concentration, pH, calcilum concentration, etc.,\cite{weiselbiophys}) under which a clot is made \cite{Per1,Per2}. In fact, \cite{Mac3} 
summarizes several reasons in addition to fibrin assembly conditions (such as, protein packing density, lateral binding between protofibrils. etc.) that 
result in the 
variation of fibrin persistence length. For our fibrin networks, which have been derived from plasma clots, the average diameter of the fibers is about 
$220$nm \cite{Oleg}, which together with a Young's modulus $E = 5$MPa \cite{WeiselPNAS} gives a persistence length ${EI}/{k_{B}T}$ at room temperature 
about 1m. The length $l$ between branch points is about $1-2\mu$m, on average. Thus, thermal fluctuations of the fibers are negligible, consistent with the 
assumption for foams.}}

When large compressive strain is applied to a fibrin network the total volume decreases considerably and the inter-fiber space is reduced. In this configuration a linear stress-strain relation is not applicable any more. The fibers are mostly buckled and forced to touch each other. We call this kind of configuration the ``densified phase" or the high-strain phase. This sort of densified network has been studied by Toll \cite{Toll} who showed that the number of contact points $N_{c}$ per unit volume is an increasing function of fiber volume fraction $\phi$ and was given by
\begin{equation} 
{N_c} = \frac{{16}}{{{\pi ^2}}}\frac{f}{{{d^3}}}{\phi ^2},
\label{Nc}
\end{equation}
where $f$ is a scalar invariant of the fiber orientation distribution function and is equal to $\pi/4$ for three dimensional isotropic networks \cite{Toll}.  Following \cite{foam}, in the densified phase we assume the fiber volume fraction $\phi = \phi_0/(1 - \varepsilon)$. 
In other words, all the volume change is due to the reduction in the thickness of the network in the direction of the compressive force with no change in the cross-sectional area. Our experiments indicate that this is a good approximation. We compared Eqn. (\ref{Nc}) to the experimental
data on contact point density $N_{c}$ from \cite{Oleg} and found that it indeed varied quadratically with the fiber volume fraction $\phi$ as shown in 
Figure~\ref{phac}. The stress-strain relation for such a network was proposed by van-Wyk \cite{Wyk} and Toll \cite{Toll} in the following form:
\begin{equation}
\sigma  = k{E_s}\left( {{\phi ^n} - \phi _0^n} \right),
\label{E2}
\end{equation}
where the coefficient $k$ is determined by the material and loading conditions. It is found to be less than $1$, as in \cite{k}. The exponent $n$ was analytically derived to be $3$ from Eqn. (\ref{Nc}) for three-dimensional random networks in \cite{Toll}. This has been confirmed by several experiments \cite{k,n1,n2,n3}. From Eqn. (\ref{E2}), we can compute the local tangent moduli of the network in the densified regime. The tangent modulus at any strain in the 
densified regime is directly proportional to the storage modulus in the rotational rheometer oscillation experiments. Following the linear theory for foams 
at small strains we assume that the ratio between this storage modulus and the local tangent modulus in the densified regime is $3/8$ (see Eqns. (\ref{E1}) 
and (\ref{G1})) and get the following expression for the shear modulus of the densified network at strain $\varepsilon$:.  
\begin{equation}
G = \frac{{9kE\phi _0^3}}{{{{8\left( {1 - \varepsilon } \right)}^4}}}.
\label{G2}
\end{equation}
Any contributions due to the Poisson effect in the densified regime are absorbed into the constant $k$ that has been treated as a fitting parameter in the 
literature. This is also consistent with our experimentally motivated assumption above that vertical compression of the fibrin network does not change its
cross-sectional area. For the plateau stress $\sigma = \sigma_0$, we find that the Young's and shear moduli are
\begin{equation}
E_2 = \frac{{3kE\phi _0^3}}{{{{\left( {1 - \varepsilon_2 } \right)}^4}}},
\label{E22}
\end{equation} 
\begin{equation}
G_2 = \frac{{9kE\phi _0^3}}{{{{8\left( {1 - \varepsilon_2 } \right)}^4}}},
\end{equation} 
where $\varepsilon_2$ is the strain at which $\sigma_0  = k{E_s}\left({{\phi ^3} - \phi _0^3} \right)$.

We will use a linearized version of the stress-strain relations (in the neighborhood of $\sigma = \sigma_{0}$) in our subsequent analysis. These are 
summarized for the low- and high-strain phases as follows.
\begin{equation}
\varepsilon \left( \sigma  \right) = \left\{ {\begin{array}{*{20}{c}}
{\sigma /{E_1}}\\
{\sigma /{E_2} + {\varepsilon _T}}
\end{array}\begin{array}{*{20}{c}}
{\sigma  \le {\sigma _0}\left( {\varepsilon  < {\varepsilon _1}} \right)}\\
{\sigma  \ge {\sigma _0}\left( {\varepsilon  > {\varepsilon _2}} \right)}
\end{array}} \right.
\end{equation}
where $\sigma_0$ is the critical stress for buckling, as discussed previously. At this stress the straight and densified phases co-exist. $E_1,E_2$ are Young's moduli in the straight and densified phases respectively from Eqns (\ref{E1}) and (\ref{E22}). $\varepsilon_T$ is the transformation strain which can be estimated by constructing the tangent to the stress-strain curve at $\varepsilon = \varepsilon_{2}$ and reading off the intercept on the $\varepsilon$ axis as shown in Figure~\ref{pha}. If we assume that the process of loading is quasi-static so that the sample stays in equilibrium during the compression process, then the compressive strain $\varepsilon$ in the plateau corresponding to the stress $\sigma_{0}$ is obtained simply by changing the fraction of fibers buckled. 
If we denote the fraction of fibrin in the high strain phase as $s \in \left[ {0,1} \right]$, then total strain can be estimated from the insight that the 
sample consists of two `springs' in series consisting of the straight and densified network respectively as observed in Figure~\ref{3d}.
\begin{equation}
\varepsilon  = \left( {1 - s} \right)\frac{\sigma }{{{E_1}}} + s\left( {\frac{\sigma }{{{E_2}}} + {\varepsilon _T}} \right).
\label{eq2}
\end{equation}
This completes the description of the stress-strain curve for a fibrin network under compression. As an example we have plotted this stress-strain curve for a fibrin network with network parameters experimentally measured  in Kim {\it et al.}~\cite{Oleg} in Figure \ref{phac}. We have used values of the fiber 
Young's modulus $E_{s}$ = 5 MPa~\cite{Science, Acta, singlefiber}. The stresses in Figure~\ref{phac} are of the same magnitude as the measured storage moduli in Figure~\ref{G}. {{However, the stress strain curve in Figure~\ref{ws} does not have an initial linear regime for small strains even though it
seems to have a flat plateau regime for intermediate strains and a regime with steeply increasing stress for high strains.}}
Furthermore, the {{shear moduli measured in Figure~\ref{G} are at least three orders of magnitude lower than those seen in the normal stress data plotted in Figure~\ref{ws}. This indicates that Figure \ref{ws} is not a true reflection of the stress strain response of the network; rather it is the result
of water squeezing out of the network in response to compression.} }
To understand why this is the case we must account for the strain-rate dependence of the mechanical response of these fibrin {{gels}}. 
 
\subsection{Strain Rate Dependence} 
\begin{figure*}[ht]
\centering
\includegraphics[width=2.3in]{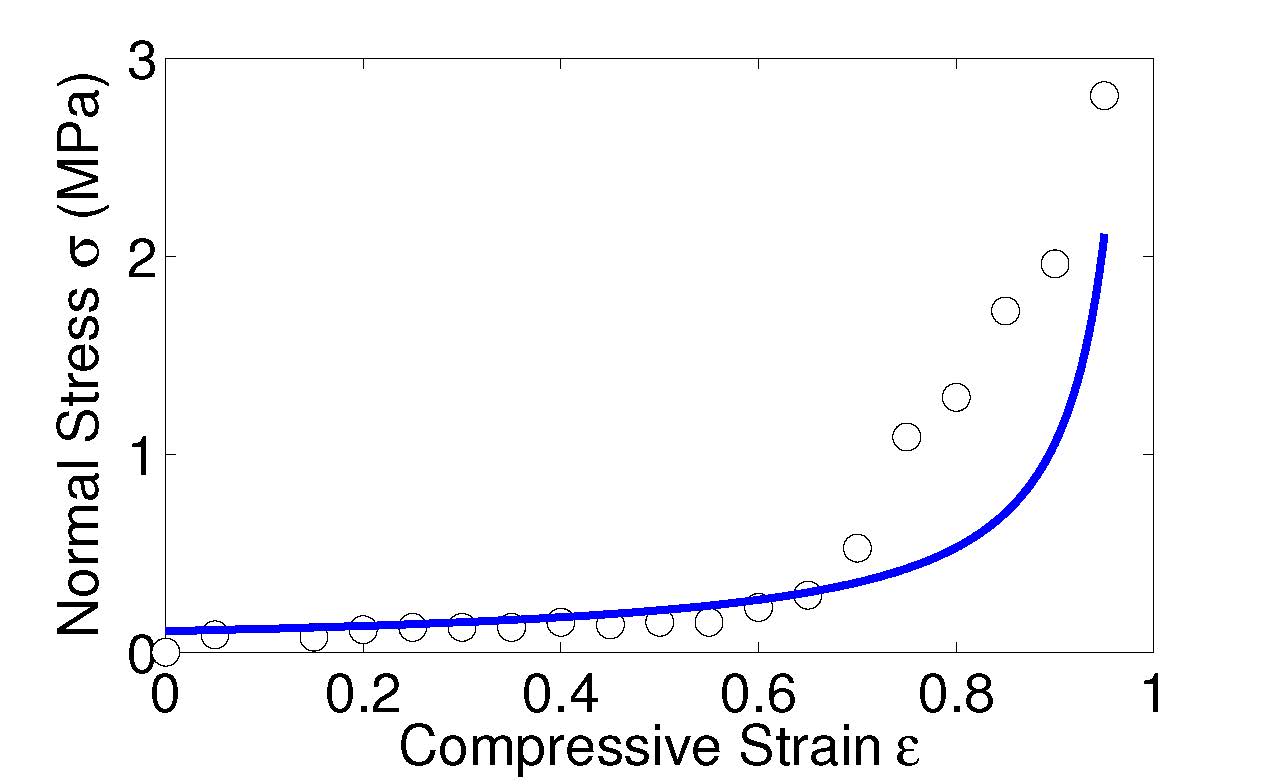} 
\includegraphics[width=2.3in]{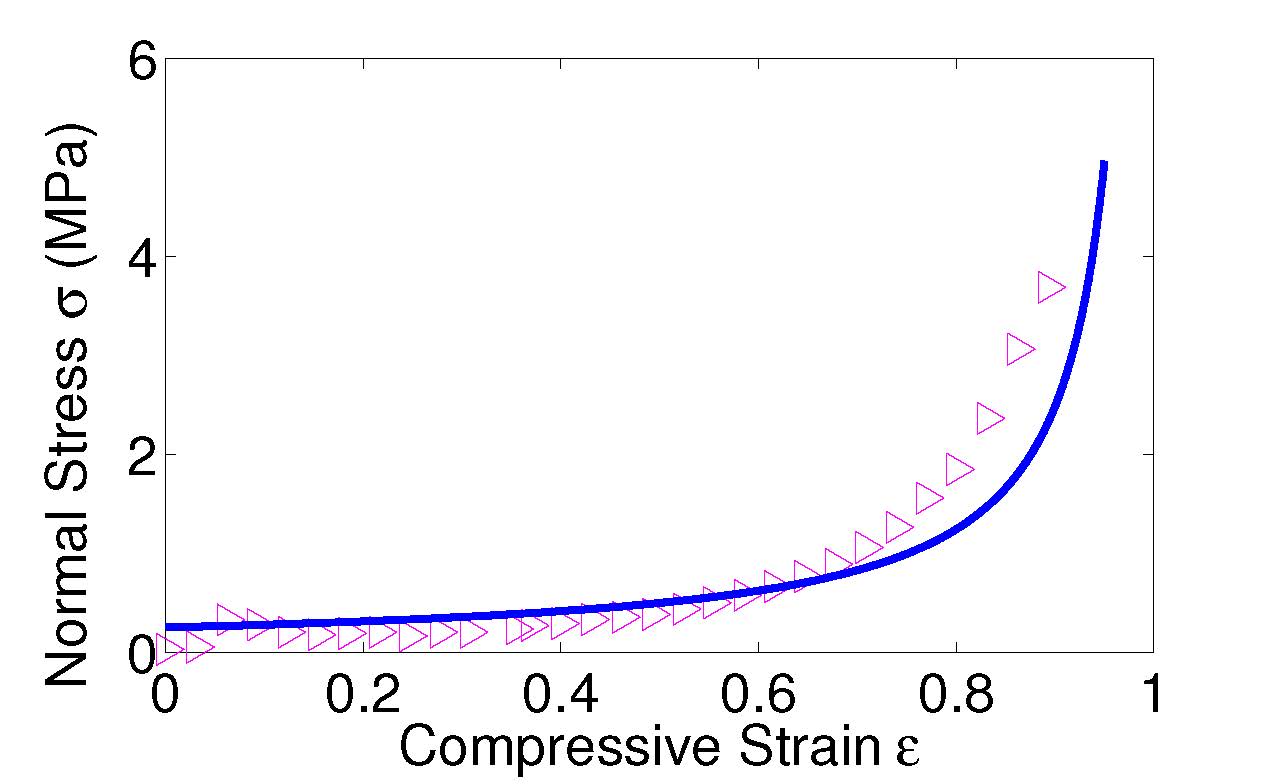} 
\includegraphics[width=2.3in]{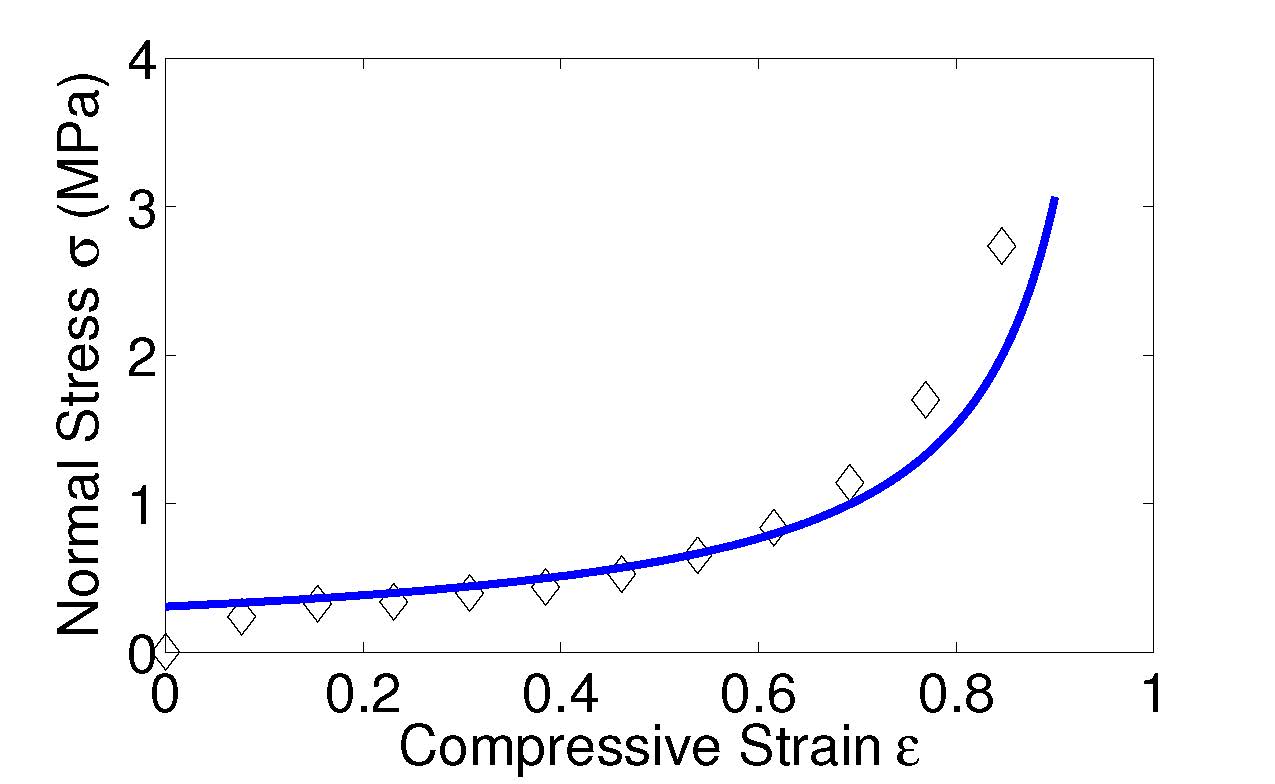} 
\includegraphics[width=2.3in]{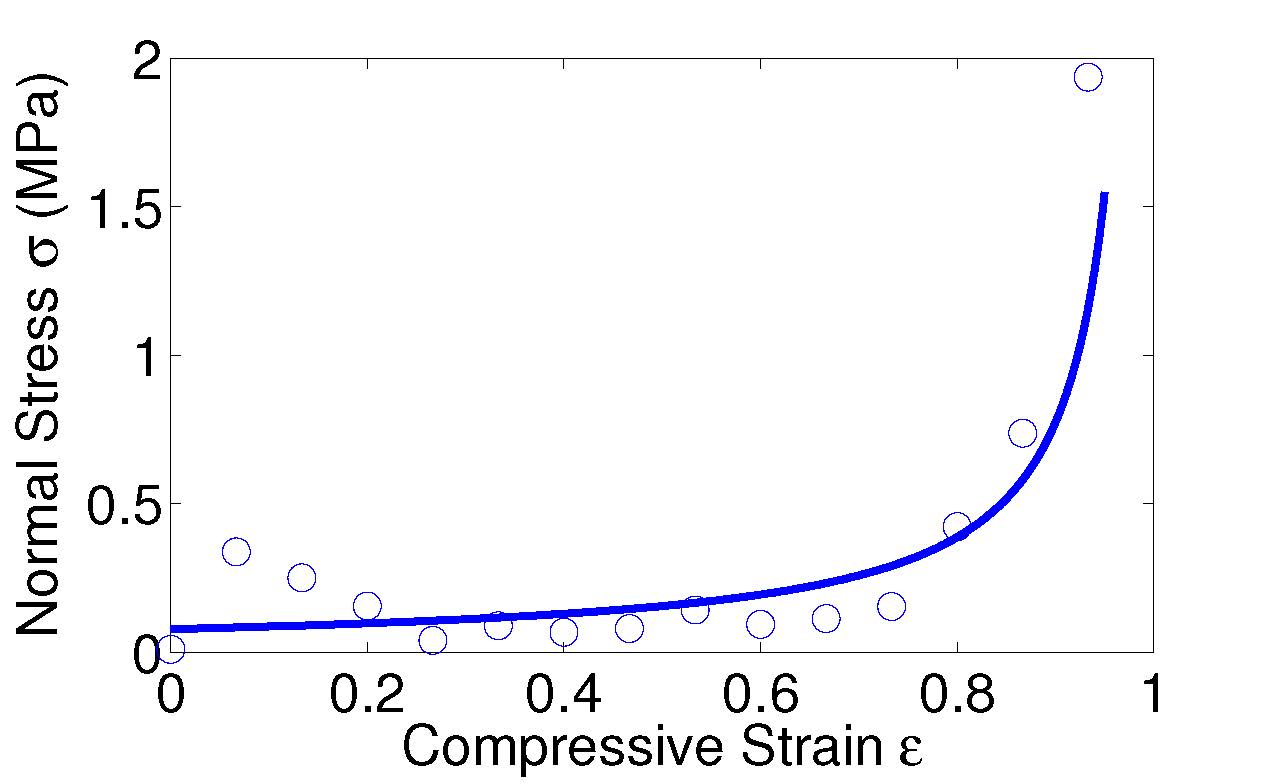} 
\includegraphics[width=2.3in]{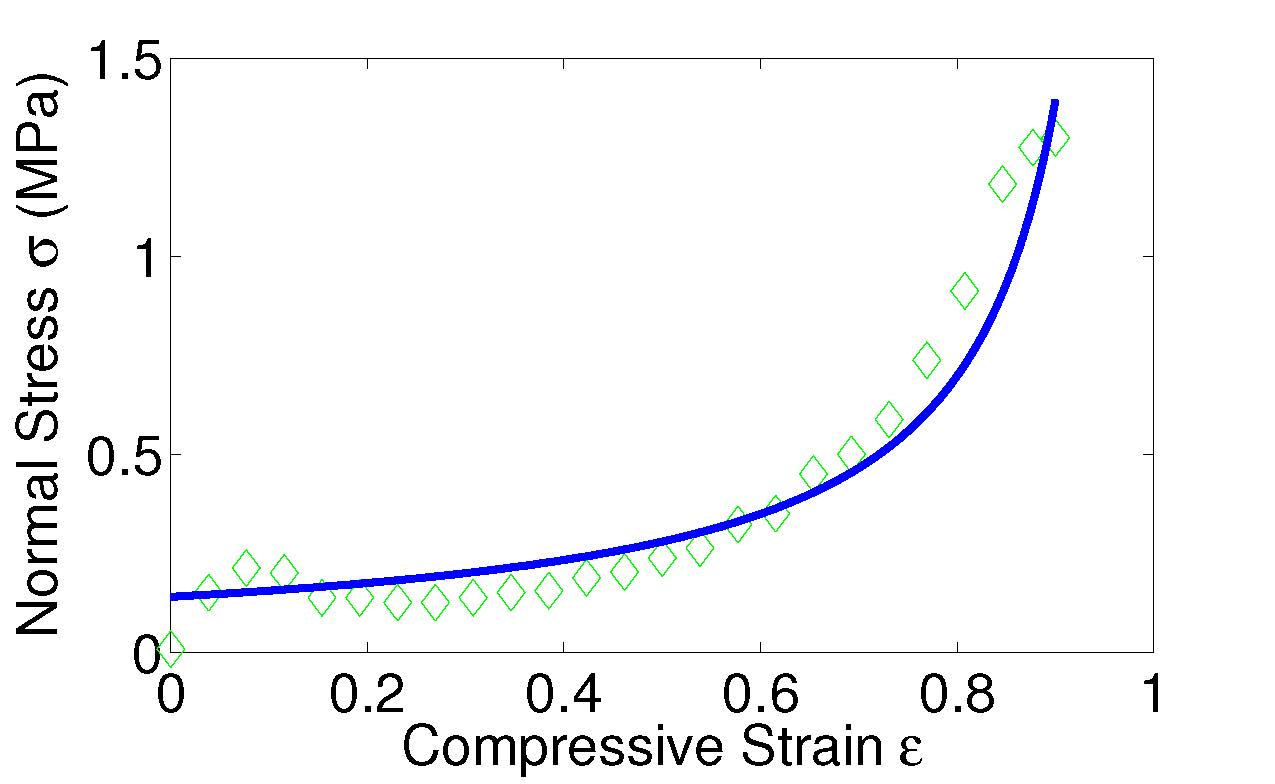}
\includegraphics[width=2.3in]{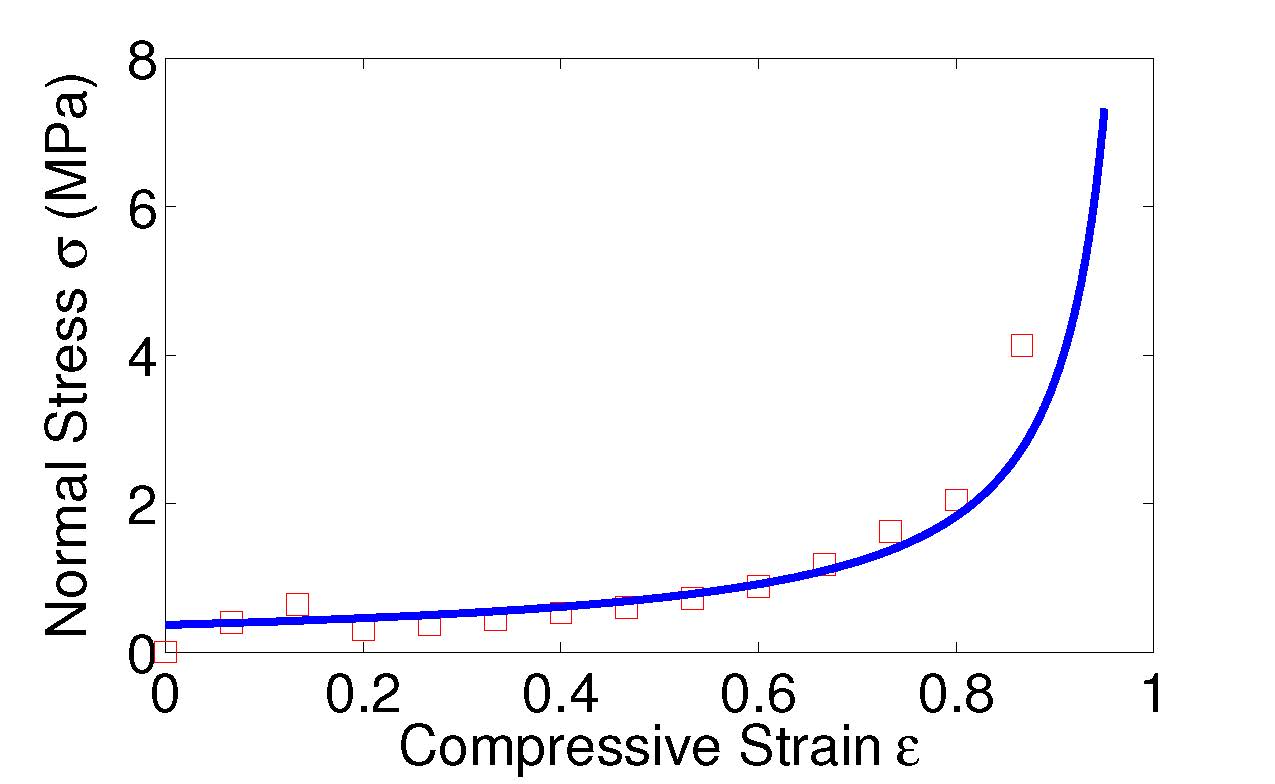} 
\caption{Fitting of the experimental {data} of normal stress according to Eqn. (\ref{cs}). Different symbols and colors are experimental data corresponding to different clot samples {{of varying height as shown in Table \ref{Fitting}. Since the rate at which they are compressed is $30\mu$m/s for all
samples the strain rate for each of them is different}}. Lines show model fits with parameters as given in Table \ref{Fitting}. Most of the normal stress 
measured in the compression experiments on our fibrin gels is due to the escape of water in response to compression. {{For this reason we cannot
extract the storage and loss modulus of the networks from these stress-strain curves.}}} 
\label{ws}
\end{figure*}

The fibrin networks that have been tested in our experiments really are gels that contain a large amount of water. As the gel is compressed, water is squeezed out. However, as it is sheared, water is not squeezed out because the total volume does not change. Hence, it is expected that the measured normal stresses in compression will depend on the applied strain rates. This effect is well-known as poro-viscoelasticity in foams and other cellular solids \cite{foam}. An analytic relation for the stress as a function of the applied strain rate is given as \cite{foam}
\begin{equation}
\sigma  = \frac{{C\mu \dot \varepsilon }}{{1 - \varepsilon }}{\left( {\frac{D}{l}} \right)^2},
\label{cs}
\end{equation}
where $\mu$ is dynamic viscosity of the fluid, $\varepsilon$ and $\dot \varepsilon$ are the compressive strain and its rate, $D$ is horizontal dimension of the foam sample, and $l$ is the cell edge-length of the foam. The coefficient $C$ is about unity. As shown in Figure~\ref{ws}, the measured normal stress during compression can be captured by Eqn. (\ref{cs}) with the parameters $\mu =0.001$ Pa$\cdot$s (for water), $D=22$ mm (known from our experimental set-up) and $l$ 
as the only fitting parameter for each experiment. {{The strain rate for each experiment is different because the height of the samples varies while the
rate of compression ($30\mu$m/s, see section 2.3) remains fixed for each experiment.}} The fitted values of $l$ appear in Table \ref{Fitting}. They are within the range of variation 
of fiber lengths {{between branch points}} seen in experiments~\cite{Oleg}. We see that most of the normal stress measured in the experiments is 
due to the expulsion of water from 
the fibrin gel. {{Hence, the storage and loss moduli measured in our rheometer experiments and shown in Figure~\ref{G} cannot be extracted from 
the stress-strain curves shown in Figure~\ref{ws}.}}

\begin{table*}[]
\caption{Fitting parameters for each experiment.}
\centering

\begin{tabular}{lp{60pt}p{60pt}p{60pt}}
	\toprule
	Individual group & Sample height(mm) &  strain rate($\text{s}^{-1}$) &  Fiber length($\mu$m) \\
	\midrule
	Black circle & 550 & 0.055 & 0.5 \\
	Magenta triangle & 650 & 0.046 & 0.3 \\
	Black diamond & 650 &0.046 & 0.27 \\
	Blue circle & 750 & 0.04 & 0.5 \\
	Green diamond & 650 &0.046 & 0.4 \\
	Red square & 750 & 0.04 & 0.3 \\
	\bottomrule
\end{tabular}

\label{Fitting}
\centering
\begin{tabular}{lp{27pt}p{27pt}p{22pt}p{57pt}p{66pt}}
	\toprule
	Individual group & $\phi_0$ &  k & $\varepsilon_{20}$ & M ($\text{Pa}^{-1}\text{s}^{-1}$) & $ w\tau_{on} (10^{-18}\text{J}\cdot  \text{s})$\\
	\midrule
	Black circle & 0.0045 & 0.025 & 0.82 & 0.004 & 2 \\
	Magenta triangle & 0.0045 & 0.0025 & 0.8 & 0.005 & 1.2 \\
	Black diamond & 0.0035 & 0.0025 & 0.85 & 0.008 & 1.7 \\
	Blue circle & 0.0035 & 0.333 & 0.7 & 0.008 & 4\\
	Green diamond & 0.004 & 0.333 & 0.6 & 0.008 & 6.5 \\
	Red square & \multicolumn{5}{c}{Densification not observerd} \\
	\bottomrule
\end{tabular}

\end{table*}

From Eqn (\ref{cs}), we can approximate the viscosity of the gel for small strains (infinitesimal values of $\varepsilon$) as
\begin{equation} \label{eq:esti} 
\eta \propto \mu{\left( {\frac{D}{l}} \right)^2}
\end{equation}
However, the viscous losses due to this term cannot be measured in the rotational rheometer experiments because the derivation of this expression assumes that the volume of the network is changing while the oscillatory experiments are performed in shear which involves no change in volume. Therefore, the loss modulus in the low strain phase should only come from the liquid viscosity itself. For water, this is merely $\mu = 0.001$ Pa$\cdot$s, which is much smaller than the value obtained from (\ref{eq:esti}). Therefore, we treat the fibrin network in the low strain phase as a purely elastic solid with constant modulus.

Another important network property that is related to the strain rate is the inter-fiber friction. This is particularly important in the densified phase in which a large number of contacts are created between fibers. Fibers sliding against each other while in contact will cause dissipation in the network. In order to estimate the viscosity associated with this process we appeal to an elegant calculation in \cite{ff}. This book gives an explanation for the molecular basis of ``viscosity" due to the forming and breaking of bonds as molecules slide past each other. The expression is:
\begin{equation} 
\eta  = w N_c  \tau_{on},
\label{eq:etahigh}
\end{equation} 
where $w$ is the inter-molecular bond energy and $\tau_{on}$ the average life-time of a bond. An expression for $N_{c}$ as a function of fibrin volume fraction is already given in Eqn. (\ref{Nc}). However, $w$ and $\tau_{on}$ are difficult to determine for fibrin fibers. Some work along this direction has been done for estimating the viscous losses in microtubule networks~\cite{mtu}, but here we treat the product $w\tau_{on}$ as a fitting parameter. In summary, we model the densified network at $\sigma = \sigma_{0}$ (for small strains) as a Kelvin-Voigt solid~\cite{Fung} in which the stress-strain relation is given as:
\begin{equation}
\sigma = E_2(\varepsilon-\varepsilon_T) + \eta \dot{\varepsilon},
\end{equation}
where the parameters $E_2$ and $\eta$ are taken from Eqns (\ref{E22}) and (\ref{eq:etahigh}). In the rotational rheometer experiments the loss 
modulus is related to this viscosity as:
\begin{equation}
G''=\eta\omega,
\label{pripri}
\end{equation}
where $\omega$ is frequency of oscillation. The loss modulus for both the low- and high-strain phases can be determined using this formula if we use $\eta=\mu=0.001 $Pa$\cdot$s (of water) for the low-strain phase and Eqn.(\ref{eq:etahigh}) for the high-strain phase. In order to estimate the loss modulus in the plateau regime of the stress-strain curve we need to account for the rate at which the fractions of the two phases evolve. This is explained in the next section.

\subsection{Phase Boundary Mobility}
In the oscillation experiments, the fraction of each phase depends on both applied strain and strain rate. As shown in Figure~\ref{pha}, the phase boundary is expected to oscillate as the applied loading oscillates. We treat this process in a one-dimensional model as follows. After pre-compressing the sample, we assume that the oscillation is also in the axial direction. This is different from the experiments in which an oscillating shear strain is applied on the pre-compressed networks. However, since the oscillating shear strains applied in the experiments are small, we can assume that they will give storage and loss moduli that are proportional to those obtained in our calculations.  

During the oscillation process the sample may not be in equilibrium and thus the stress may not be $\sigma_{0}$. There will be oscillation in stress, phase fractions, as well as strain. We assume that the fraction $s$ is determined by the kinetics of transformation as in \cite{BookPhase}. This means that
the rate of change of each fraction is given as a function of stress $\sigma$. To simplify the problem we consider the particular kinetic equation \cite{Raj}:
\begin{equation}
\dot{s} = \Phi \left( \sigma  \right) = M\left( {\sigma  - {\sigma _0}} \right),
\label{eq3}
\end{equation}
or,
\begin{equation}
s-s_0 = \int_0^t M(\sigma-\sigma_0)\text{d}t = { M\int_0^t (\sigma-\sigma_0)\text{d}t  },
\label{ss0}
\end{equation}
where $M$ is a { time-independent} mobility parameter determined by the network that could be regarded as a fitting parameter in {{our}} theory and $s_0$ is the fraction of high strain phase before oscillation is applied. In the experiment we apply a harmonic oscillating strain to the pre-compressed network as:
\begin{equation}
\varepsilon  = \varepsilon_0 + a\sin \omega t,
\label{eq4}
\end{equation}
where $\varepsilon_0 $ is an initial strain in the plateau regime of phase transition before oscillation is applied. Taking account of the strain rate
dependence we rewrite Eqn. (\ref{eq2}) as:
\begin{equation} 
\begin{split}
\varepsilon &=(1-s)\varepsilon_{1} + s \varepsilon_{2}, \\
\sigma &= E_1\varepsilon_{1} = E_2(\varepsilon_{2}-\varepsilon_T) + \eta \dot{\varepsilon_2}. 
\end{split}
\label{eq:strfrac}
\end{equation}
Now, recall that the expression for equilibrium stress $\sigma_0$ and strain $\varepsilon_0$ are given by:
\begin{equation}
\begin{split}
\varepsilon_0 &=(1-s_0)\varepsilon_{10} + s_0 \varepsilon_{20}, \\
\sigma_0 &= E_1\varepsilon_{10} = E_2(\varepsilon_{20}-\varepsilon_T). 
\end{split}
\label{stressstrain0}
\end{equation}
where the subscript $0$ denotes the quantity in initial compression equilibrium before oscillation. Subtracting Eqns. (\ref{stressstrain0}) from (\ref{eq:strfrac}) we get:
\begin{equation}
a \sin \omega t = (1-s_0)x + s_0y+(s-s_0)(\varepsilon_2 - \varepsilon_1),
\label{governing}
\end{equation} 
\begin{equation}
\sigma - \sigma_0 = E_1 x = E_2 y + \eta \dot{y},
\label{sigmasigma0}
\end{equation}
where $x=\varepsilon_1 - \varepsilon_{10}$ and $y=\varepsilon_2 - \varepsilon_{20}$ are the small oscillating components of strain in each phase. Since the 
oscillation in strain is small, we make an approximation in Eqn. (\ref{governing}) by replacing $\varepsilon_2 - \varepsilon_1$ with 
$\varepsilon_{20} - \varepsilon_{10}$. Then, plugging Eqn. (\ref{ss0}) and (\ref{sigmasigma0}) in (\ref{governing}) we get:
\begin{equation}
\begin{split}
a \sin \omega t &= (1-{s_0})\dfrac{E_2y+\eta \dot{y}}{E_1} + {s_0}y \\
&+ M(\varepsilon_{20} - \varepsilon_{10}) \int_0^t (E_2y+\eta\dot{y})\text{d}t.
\end{split}
\end{equation}
Carrying out the integral of $\dot{y}$ and rearranging terms, we find that
\begin{equation}
a \sin \omega t =  Ay+B\dot{y}+C\int_0^t y \text{d}t,
\label{7}
\end{equation}
where the three dimensionless quantities $A$, $B$, and $C$ are given by:
\begin{equation}
\begin{split}
A &= (1-{s_0})\dfrac{E_2}{E_1} + {s_0}+M\eta(\varepsilon_{20} - \varepsilon_{10}),\\
B &= (1-{s_0})\dfrac{\eta}{E_1}, \\
C &= ME_2(\varepsilon_{20} - \varepsilon_{10}).
\end{split}
\end{equation}
Let us assume that this equation has the solution: 
\begin{equation}
y =U\sin\omega t + V\cos \omega t,
\end{equation}
where $U$ and $V$ are at present unknown. Plugging this into Eqn. (\ref{7}) and solving for the two coefficients $U$ and $V$ gives: 
\begin{equation}
\begin{split}
&U = a\dfrac{A}{A^2+(B\omega-C/\omega)^2}, \\
&V = - a\dfrac{B\omega-C/\omega}{A^2+(B\omega-C/\omega)^2}.
\end{split}
\end{equation}
Plugging this back into Eqn. (\ref{eq:strfrac}) we get the expression for stress as:
\begin{equation}
\begin{split}
\sigma &= \sigma_0 + E_2 y + \eta \dot{y}\\
&= \sigma_0 + a(E_2U-\eta V\omega) \sin \omega t + a(E_2 V +\eta U\omega) \cos \omega t.
\end{split}
\end{equation}
Therefore, the storage and loss modulus in the plateau regime (i.e. $\sigma = \sigma_{0}$) are respectively
\begin{equation}
\begin{split}
G' &= {E_2}U - \eta V\omega  = \frac{{{E_2}A + \eta B{\omega ^2} - \eta C}}{{{A^2} + {{\left( {B\omega  - C/\omega } \right)}^2}}}, \\
G'' &= {E_2}V + \eta U\omega  = \frac{{ - {E_2}\left( {B\omega  - C/\omega } \right) + \eta A\omega }}{{{A^2} + {{\left( {B\omega  - C/\omega } \right)}^2}}}.
\end{split}
\label{FT}
\end{equation}

\begin{figure*}[ht]
\centering

\includegraphics[width=2.2in]{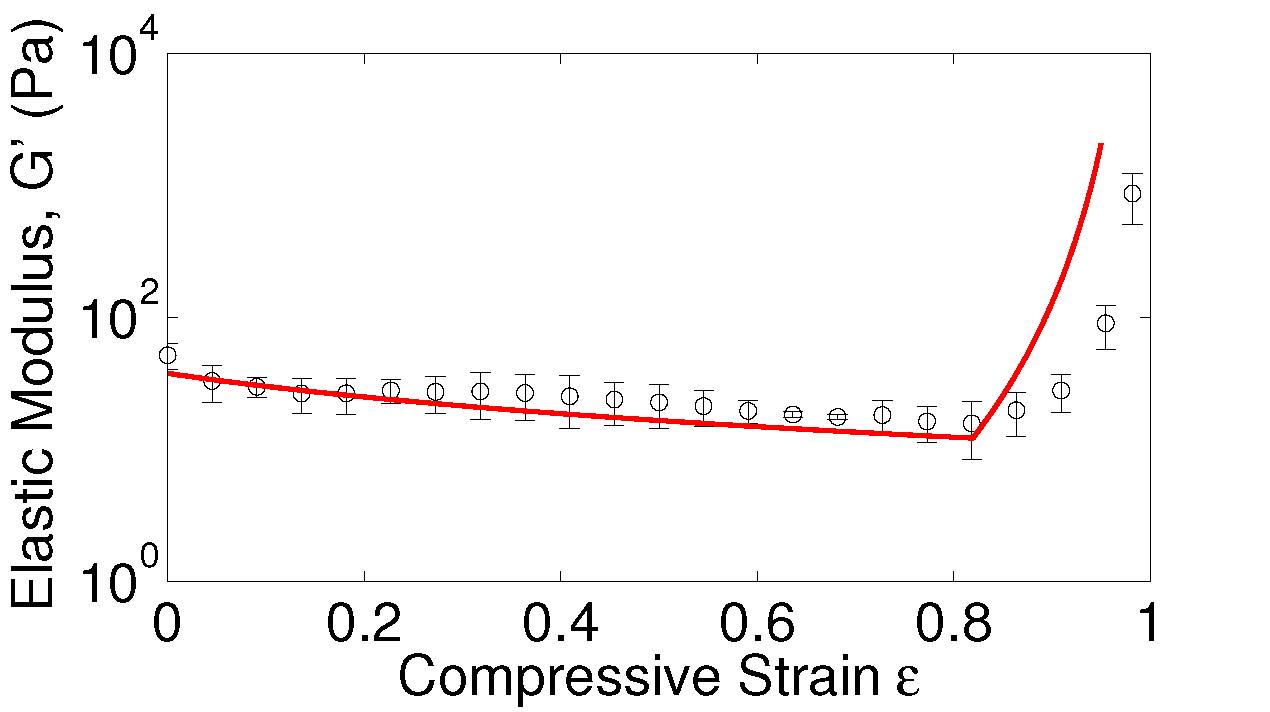} 
\includegraphics[width=2.2in]{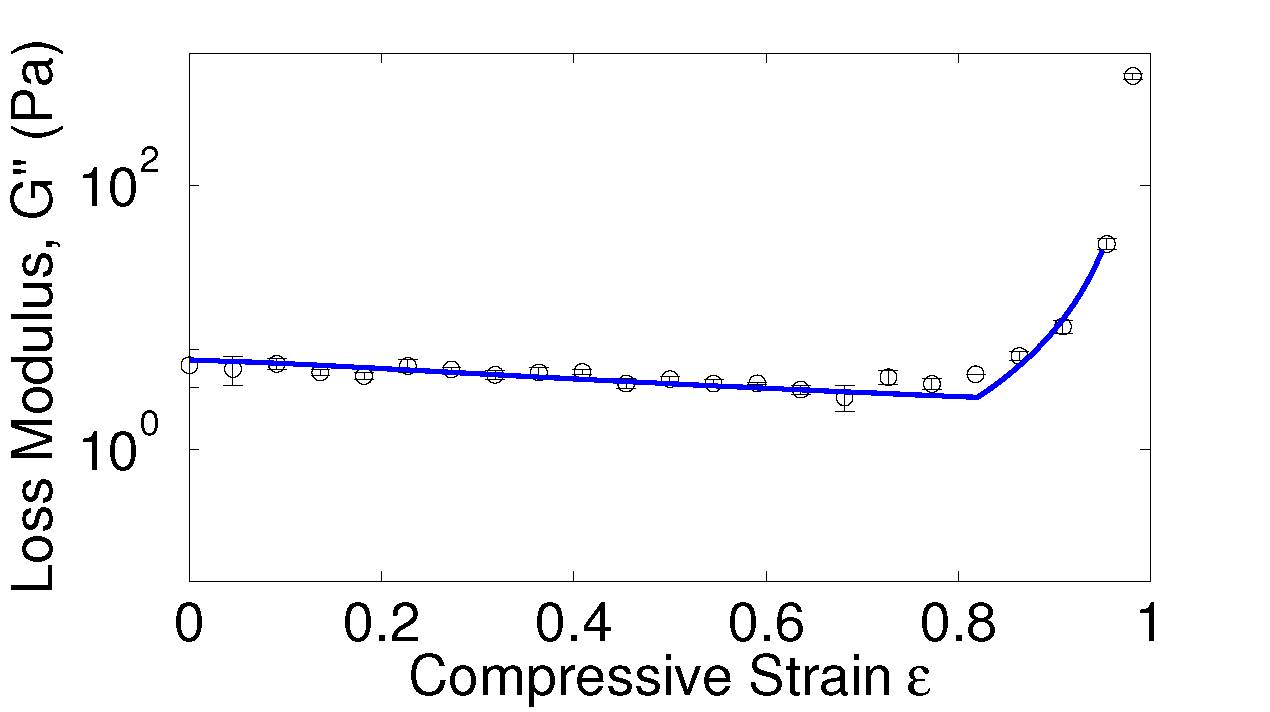} 
\includegraphics[width=2.2in]{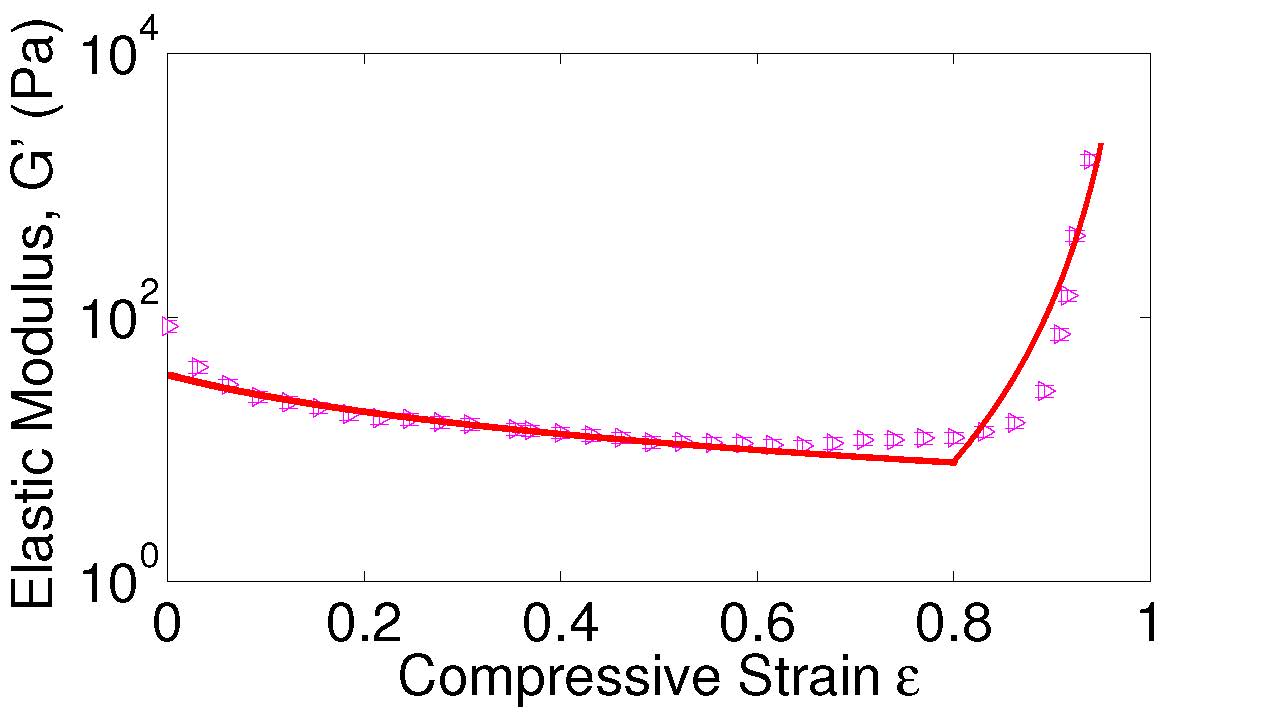} 
\includegraphics[width=2.2in]{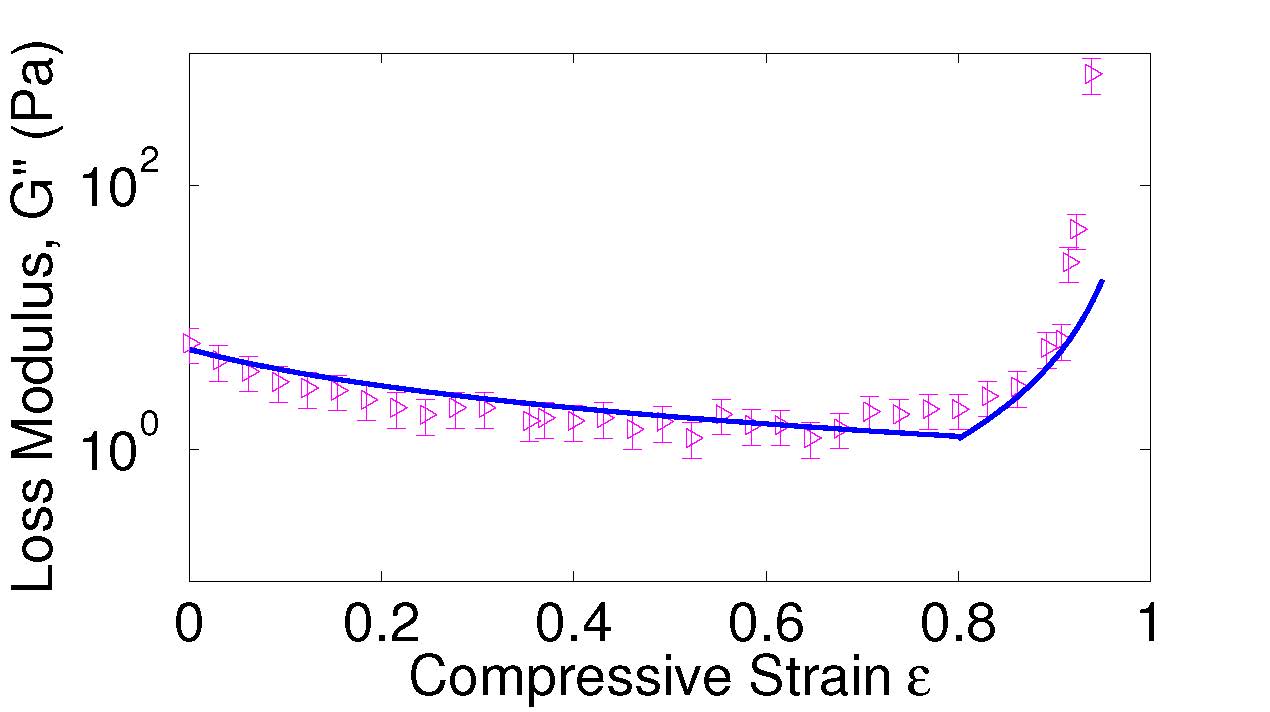} 
\includegraphics[width=2.2in]{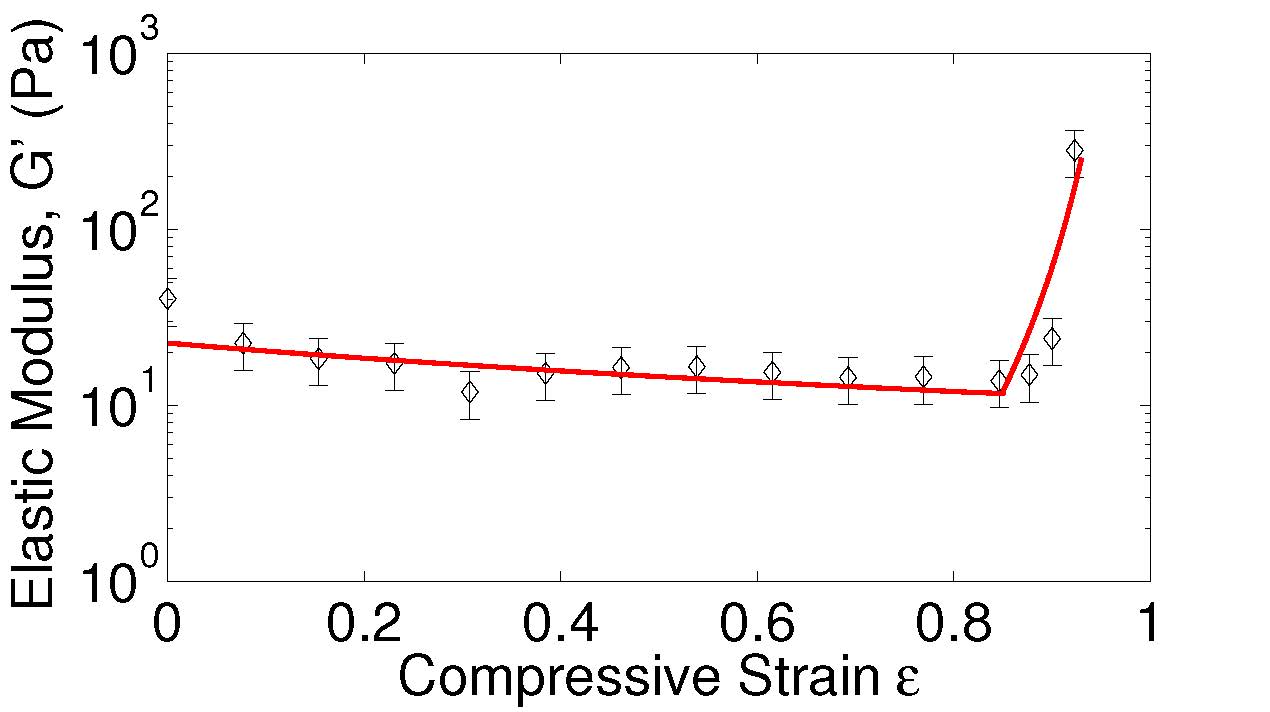} 
\includegraphics[width=2.2in]{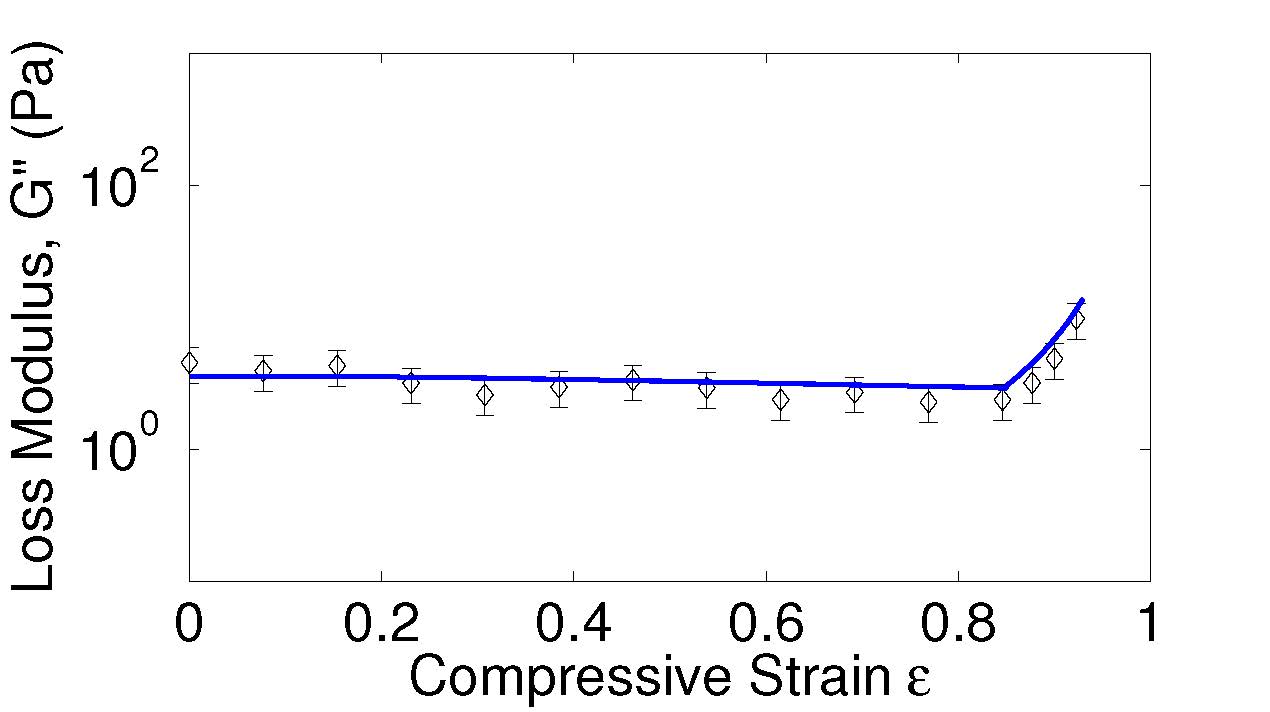} 
\includegraphics[width=2.2in]{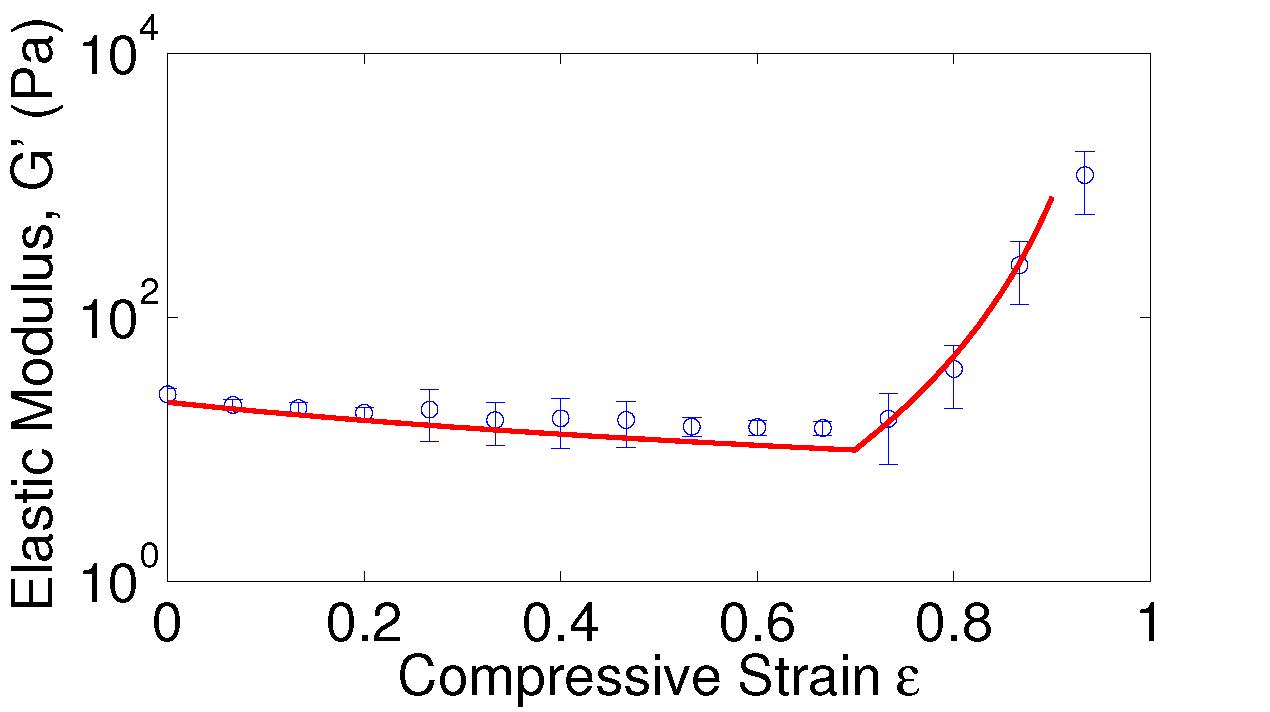} 
\includegraphics[width=2.2in]{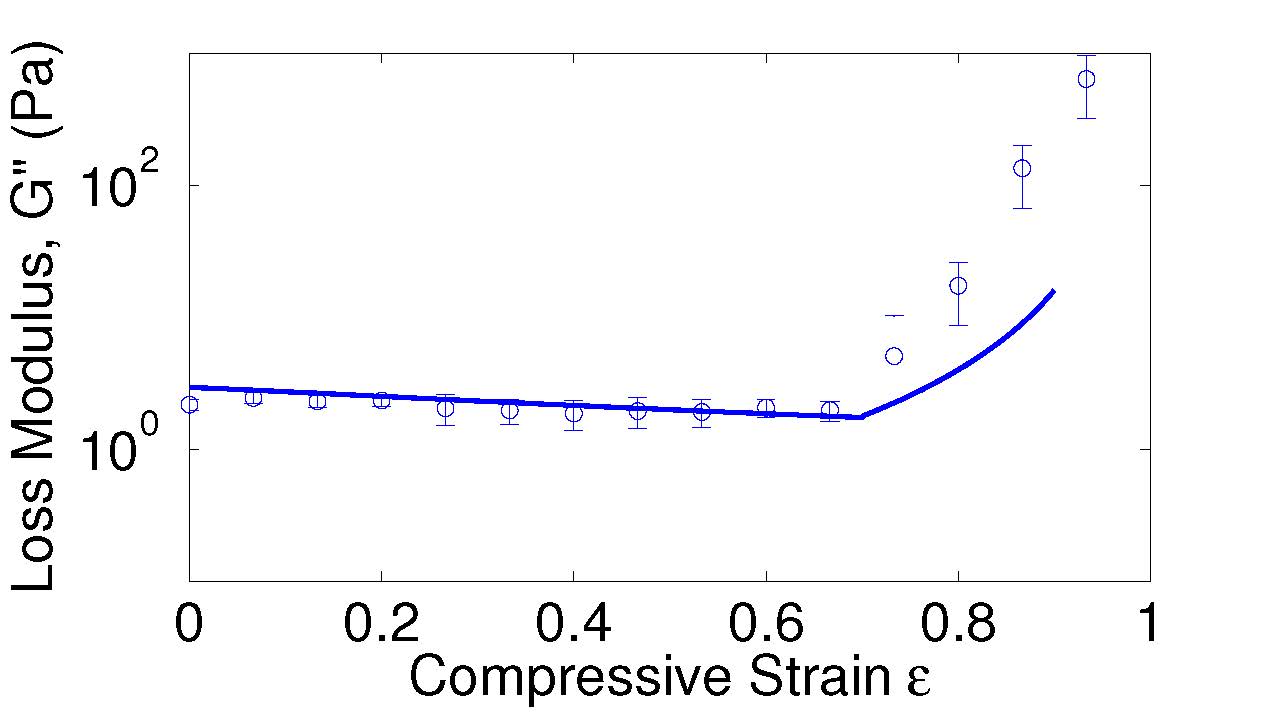} 
\includegraphics[width=2.2in]{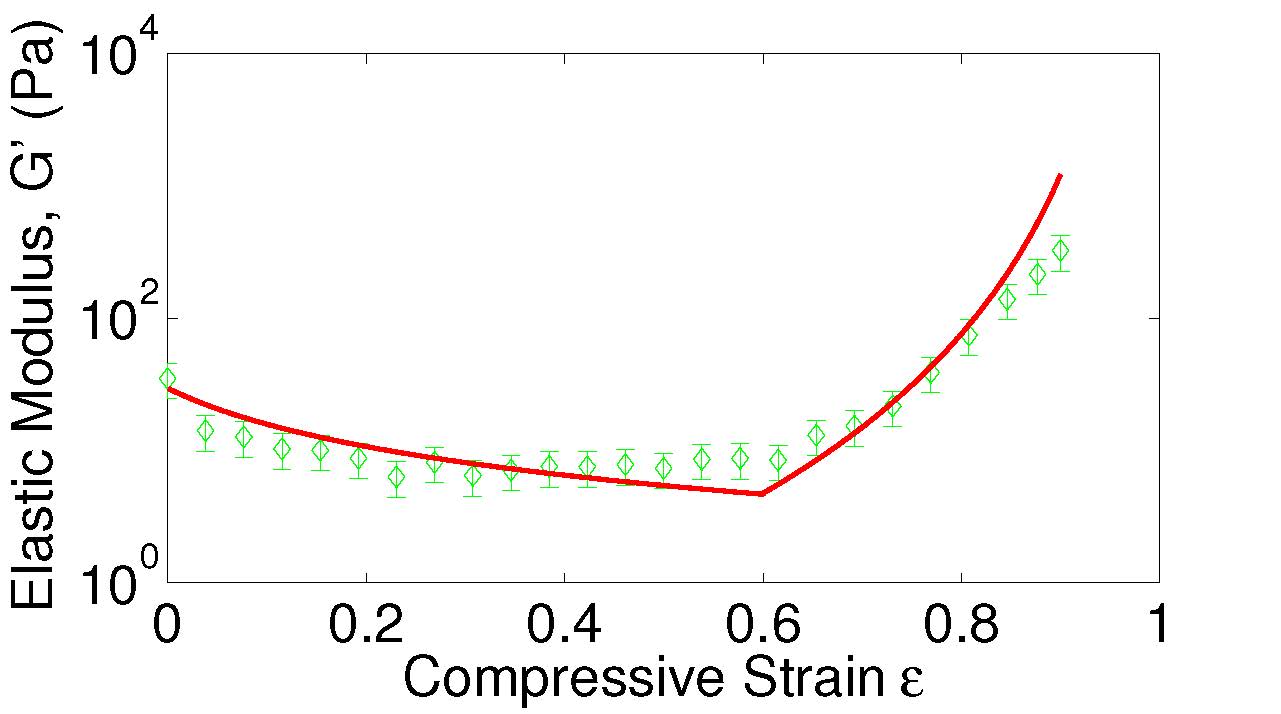} 
\includegraphics[width=2.2in]{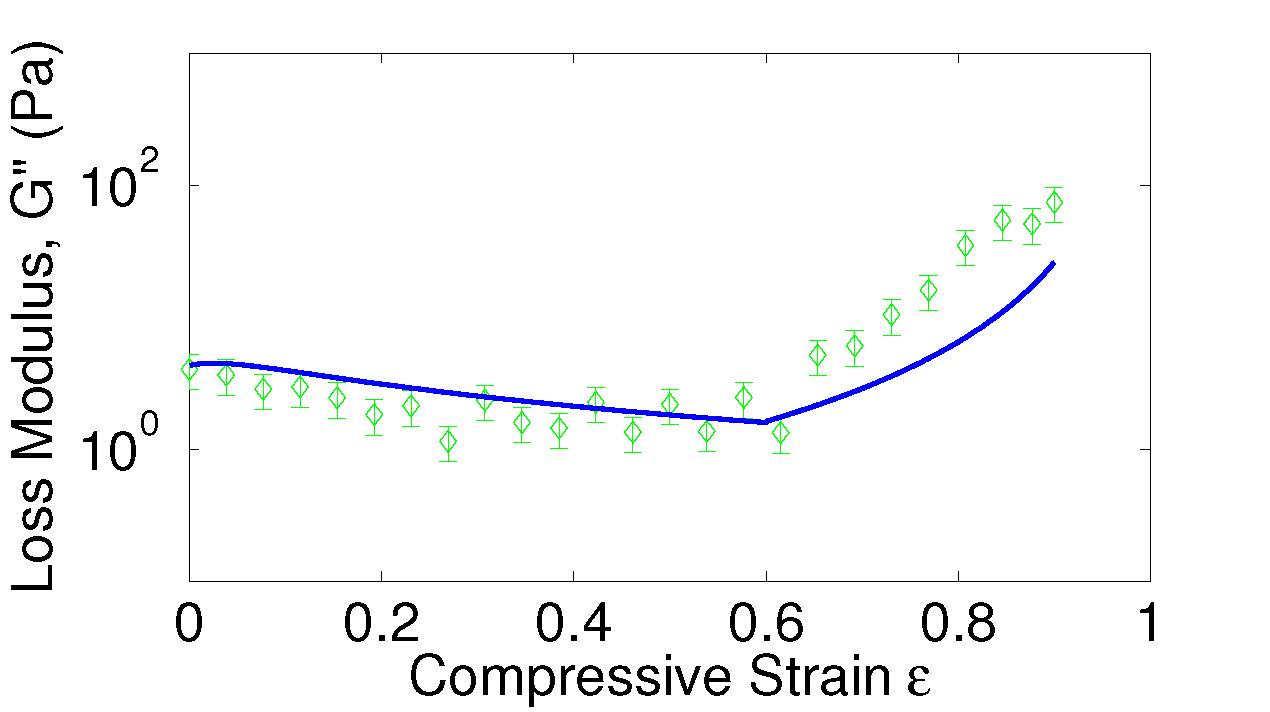} 
\includegraphics[width=2.2in]{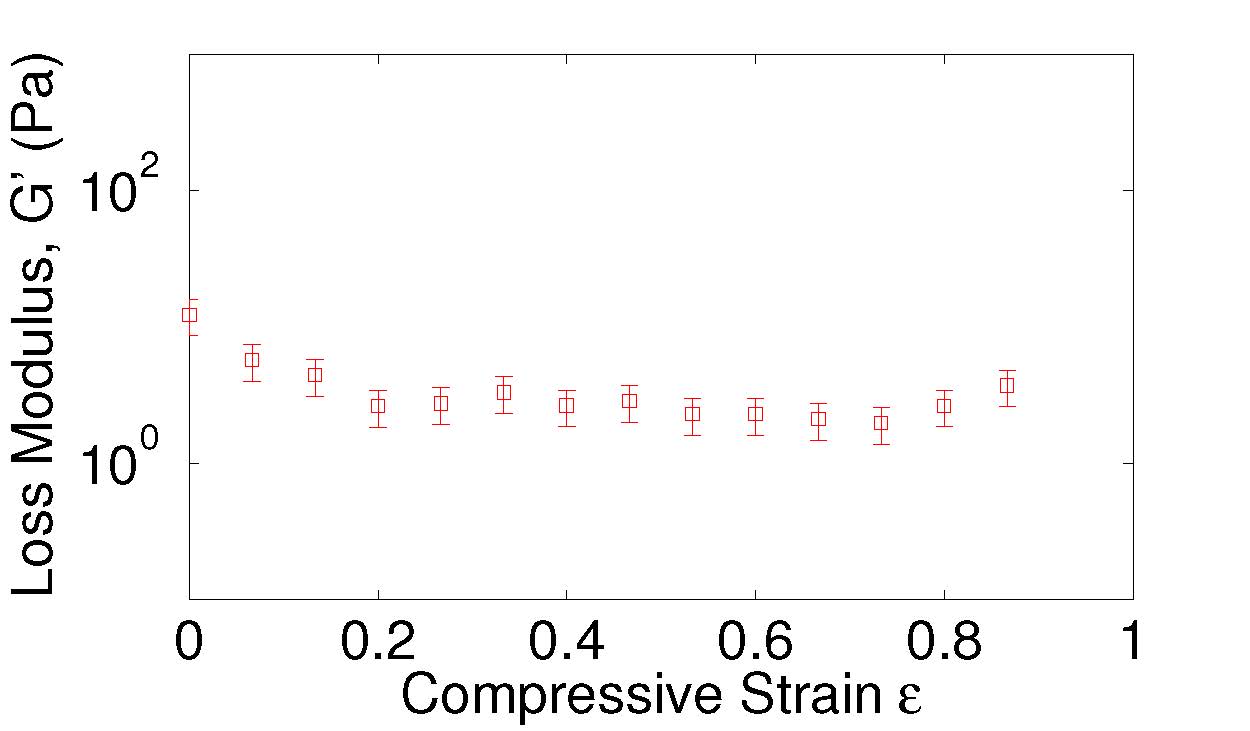}
\includegraphics[width=2.2in]{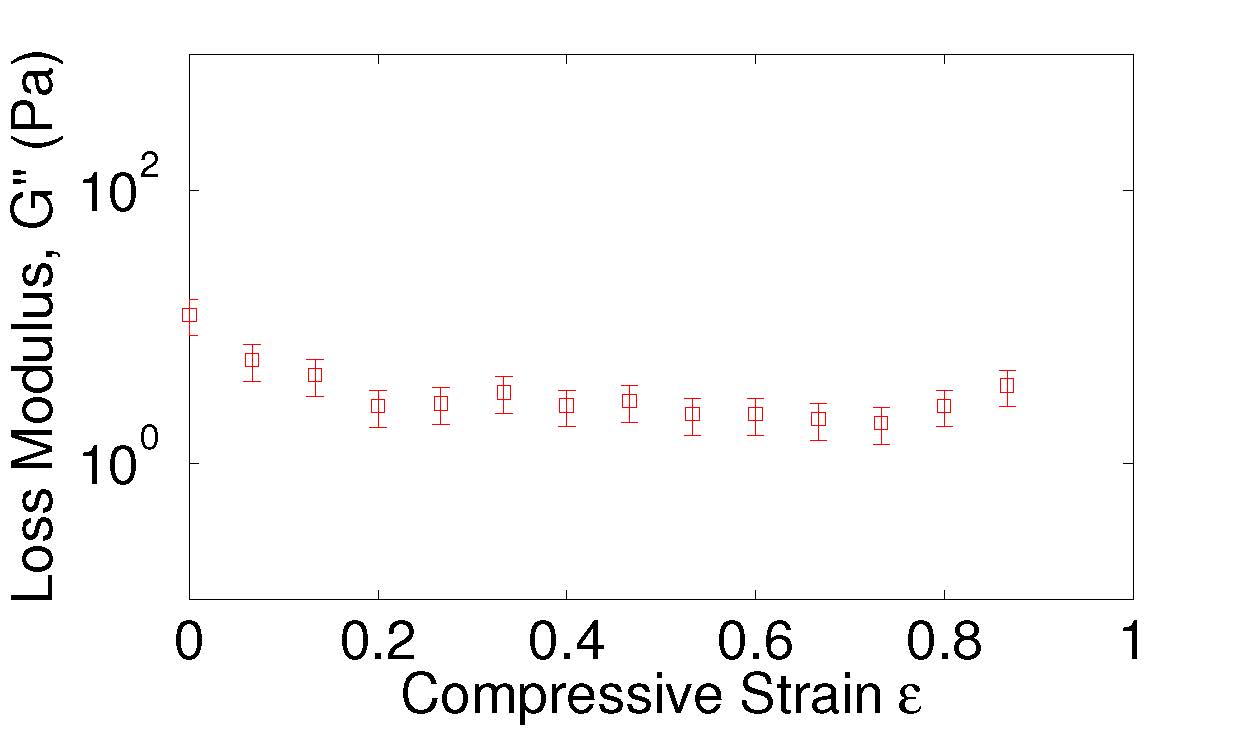} 

\caption{Fitting of the storage moduli and loss moduli according to Eqns. (\ref{G1}), (\ref{G2}), (\ref{FT}), (\ref{eq:etahigh}), (\ref{pripri}), and (\ref{FT}). We have used fitting parameters for each individual experimental group as given in Table \ref{Fitting}. {{Note that there is no densification regime in the last group. One possible reason could be the failure or damage of the network under large compression. However, the data for strains smaller
than 0.8 in the last group are consistent with the other data sets.}}}
\label{G}
\end{figure*}

In order to confirm that the above framework predicts the correct storage and 
loss moduli, we compared quantitatively our theory with the experimental data. The input parameters of the network in the above computations were taken from the experiments shown in Figure~\ref{G}. Combining Eqns. (\ref{G1}), (\ref{G2}) and (\ref{FT}) together, we obtain the theoretical curve in each individual experiment in Figure~\ref{G}. We take the value of branch point density and fiber density exactly as measured in experiment as $0.04 \mu m^{-3}$ and $0.1 \mu m^{-3}$ respectively. The other fitting parameters for each experiment are reported in Table \ref{Fitting}. The trends in storage and loss moduli are captured quite well with this model. Note that the experimentally measured average fiber length and average fiber diameter are $1.32$ $\mu$m and $220$ nm respectively~\cite{Oleg}, which result in a fiber volume fraction of $\phi_{0} = \pi \nu d^2 l /4 = 0.005$. The $\phi_{0}$ values obtained by our fits are slightly below this value but well within the range of variation of $\phi_{0}$ in the experiments~\cite{Oleg}. Also note that the value of constant $k$ in the densifed phase varies inversely with the strain $\varepsilon_{20}$ where the phase transition finished. We have not yet been able to rationalize this. 

\begin{figure}[ht]
\centering
\includegraphics[width=3.5in]{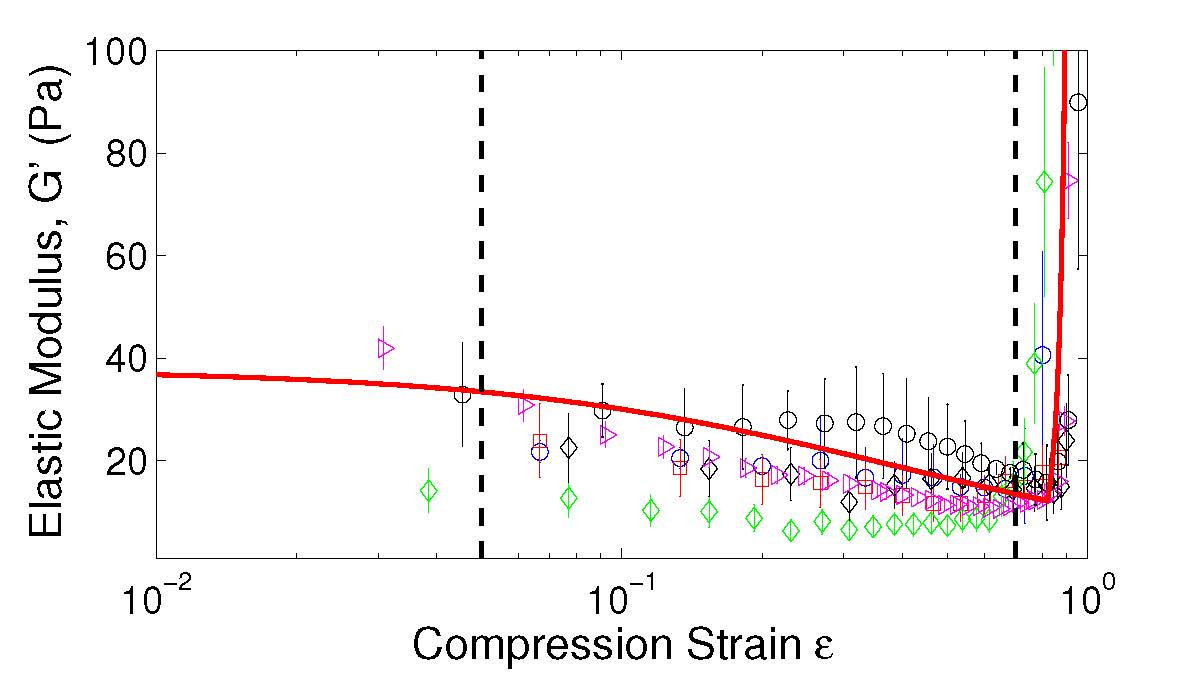}
\includegraphics[width=3.5in]{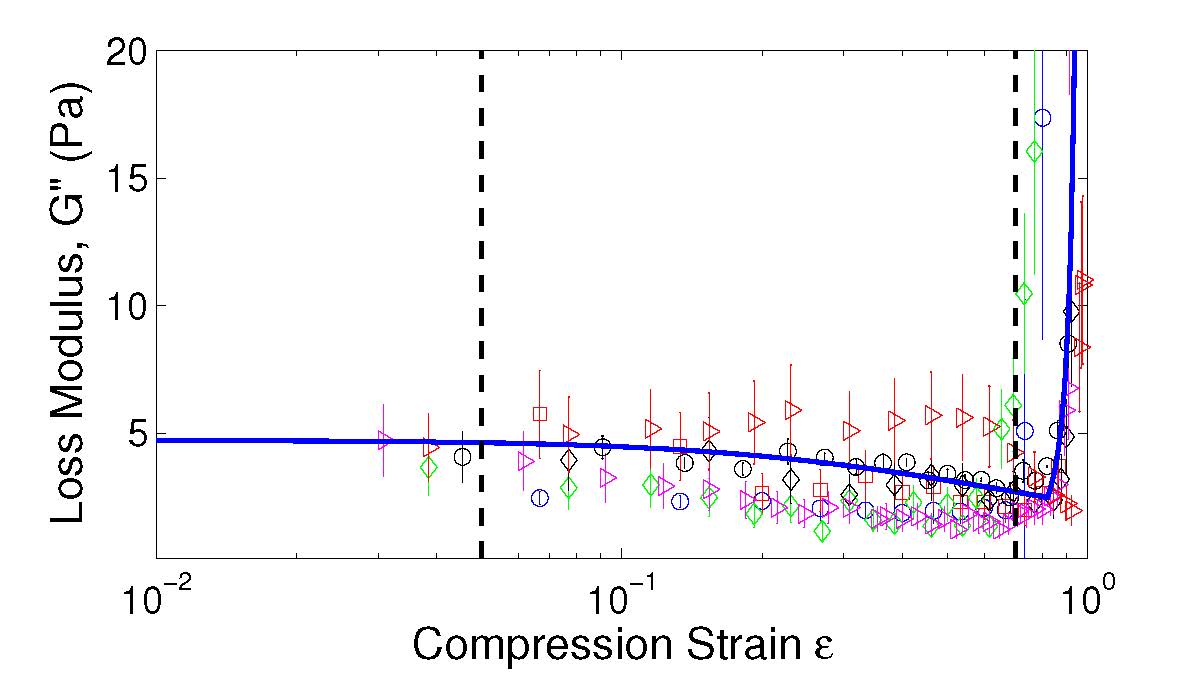}
\caption{Experimental data of storage moduli and loss moduli versus logarithm of strain, showing three different regimes of fibrin network mechanics as indicated schematically in Figure \ref{pha}. In the linear regime($0<\varepsilon<0.05$), the moduli are constant; in the phase transition regime($0.05<\varepsilon<0.7 $), they are decreasing due to progressive fiber buckling; in the desification regime ($0.7<\varepsilon<1$), they are steeply increasing due to bending of the buckled 
fibers and increased inter-fiber contact. The theoretical curves passing through the experimental data are plotted using Eqns. (\ref{G1}), (\ref{G2}), (\ref{FT}), (\ref{eq:etahigh}), (\ref{pripri}), and (\ref{FT}) with $\phi_0 = 0.0045$, $k=0.025$, $\varepsilon_{20} = 0.7$, $M = 0.004\text{Pa}^{-1}\text{s}^{-1}$, $\omega \tau_{on} = 2  \times 10^{-18} \text{J} \cdot \text{s}$.}
\label{Gini}
\end{figure}

{{We will now connect our results for the variation in moduli to Figure \ref{pha} which schematically shows three distinct regimes in the stress-strain curve. 
To this end we have replotted all the data for storage and loss moduli in Figure \ref{Gini} against the logarithm of the strain $\varepsilon$. The red and
blue curves in the upper and lower panels are fits to the average of all the storage and loss moduli, respectively. The strains $\varepsilon_{1}$ and
$\varepsilon_{2}$ have been indicated by dashed lines to demarcate the three regimes. In the initial linear regime, $\varepsilon < \varepsilon_{1}$, the 
moduli $G'$ and $G''$ are constant. In the plateau regime, $\varepsilon_{1} \leq \varepsilon \leq \varepsilon_{2}$ the network accommodates more 
strain by changing the fractions of the straight and densified phases of the networks through the motion of the phase boundary. $G'$ decreases with 
increasing strain in the plateau regime because the slope of the stress-strain curve of the densified phase at $\varepsilon = \varepsilon_{2}$ is smaller 
than the slope for the straight phase at $\varepsilon = \varepsilon_{1}$. For $\varepsilon > \varepsilon_{2}$ densification occurs and both moduli increase 
steeply.}}

\section{DISCUSSION}
We have shown in this paper that fibrin networks are natural cellular materials. The theoretical description of the fibrin mechanical behavior is novel and treats fibrin networks in the context of a  broad class of natural and synthetic foam-like materials which share some key features. Among these is the tri-phasic stress-strain response
with a rarefied low-strain phase, a densified high-strain phase, and a plateau phase consisting of a mixture of these two. Like most foams we also see
a moving phase boundary separating the two phases when fibrin networks are compressed~\cite{PB1,PB2}. The phase boundary and several other features 
quantifying the 
structural non-uniformity of fibrin networks under compression were revealed by our 3D microscopy studies. Because of the complexity of the analysis, non-uniform fibrin networks under loads have not been previously quantified. However, it is clear that under physiological conditions fibrin networks are frequently exposed to dynamic forces, which would result in non-isotropy and structural heterogeneities. Since in many cases, fibrin networks participate in the permeation of blood clotting factors \cite{zz}, variations in fibrin matrix density can affect the spatial-temporal distribution of factors and hence, alter their biochemical interaction. Our results indicate that even a uniform compression can produce structural gradients in fibrin networks, therefore, suggesting a non-uniform alteration of fibrin matrix transport properties with compression. 

It is also worth noting that the mechanical response of fibrin networks under compression is dominated by fiber {{buckling and subsequent bending}} and is therefore non-affine. This is 
in contrast to the stress-strain response in tension and shear where the assumption of affine deformations worked quite well~\cite{Science, Acta} . This assumption is particularly good for rubbers and other polymeric materials \cite{Eightchain,Flory}, and it works for fibrin networks under tension because the unfolding of the proteins results in cross-linked polymer chains much like those in rubber. It works well for shear of a network of thin fibrin fibers because such a network is not very different from cross-linked fluctuating polymer chains \cite{Mac1,Mac3}. However, in compression the initial linear regime is governed by fiber bending
instead of fiber stretching. Thus, we can no longer assume affine deformations. This is apparent in the non-linear dependence of the initial Young's modulus in compression 
$E_{1}$ on the fibrin volume fraction $\phi$ (see Eqn. (\ref{E1})). The dependence would be linear if the deformations were affine as in tension 
and shear. Since $\phi^{2} = \nu^{2} d^{4}l^{2}$ our model predicts that $E_{1} \propto \nu^{2}d^{4}l^{2}$ where $\nu$ is fiber density, $d$ is average fiber
diameter and $l$ is average fiber length. Our prediction can be tested in experiments since $\nu$, $d$ and $l$ can be controlled by changing the 
polymerization conditions of the fibrin networks. In the densified network buckled fibers are in contact, so the assumption of affine deformations is not 
expected to work. Indeed, the stress-strain response depends on $\phi$ in a complex way that has been quantified both with experiments and theory in the 
textile literature~\cite{Wyk,Toll}. We have shown in this paper that the same ideas can quantitatively capture the compression response of densified fibrin 
networks with an appropriate choice of fiber geometrical and material parameters. For example, at large compressive strains our model predicts 
that the tangent modulus in the densified phase $E_{2} \propto {\phi_{0}^{3}}/{(1 - \varepsilon)^{4}}$ where $\phi_{0}$ is the initial fiber volume 
fraction. This prediction can be tested since $\nu$, $d$, $l$ and $\varepsilon$ can all be controlled in experiments. Similarly, our physically 
motivated constitutive law Eqns. (\ref{eq:etahigh}) and  (\ref{Nc}) that accounts for the inter-fiber friction and captures the viscous response of densified 
fibrin networks can also be tested in experiments. We have already demonstrated the importance of the strain-rate dependence in the compressive response of fibrin networks by using poroelastic constitutive laws that govern the flow of water in cellular solids.   

In this work the structural alterations in response to compression were studied at the microscopic level of a network. We have shown that structural 
non-uniformity manifests itself as a ``compression front" or ``phase boundary" separating regions in the network where fibers are straight from another regime where the fibers are buckled. While such a front has been visualized in many macroscopic~\cite{PB1, PB2} and microscopic foams~\cite{Greer} we are
not aware of attempts where its motion is treated using a theory of phase transitions~\cite{Raj} to obtain the storage and loss moduli. We have demonstrated
in this work that a continuum theory of phase transitions can predict the rheological
properties of fibrin networks under large deformations if we properly account for the dissipative motion of the phase boundary. {{It is likely that
a similar model can be applied to other foams or foam-like materials since propagating interfaces have been visualized in them.}} The idea of a phase transition was also used in our earlier work on the tensile behavior of fibrin networks~\cite{Science, Acta}. The difference here is that we have no evidence of forced molecular unfolding of the protein. However, fibrin has a complex multidimensional structure and it has been shown that compression of a hydrated fibrin clot is accompanied by changes in the $\alpha$-helical coiled-coils, namely the transition of $\alpha$-helix to $\beta$-sheet revealed using FTIR spectroscopy \cite{Lin}. However, a comparative analysis shows that the changes in the secondary structure of fibrin were observed at compressive stresses ($\geq 25MPa$) that are orders of magnitude higher than those applied in our work (Figure~\ref{ws}). It is likely that molecular structural transitions observed in a protein at very high degrees of compression results from its mechanical destruction, not reached in our present experiments. The aggregate of data suggests that at the degrees of deformation, not leading to complete disorganization of a filamentous network, compressive load is accommodated mainly by the structural changes in fiber arrangements at the network level. These changes include buckling, stretching, and formation of oblique contacts of individual fibers as well as overall network densification. It is noteworthy that the structural changes underlying compressive deformation of the networks are spatially non-uniform and form a stepped gradient along the axis of compression. Although we observed a single compressive front in our experiments that originated at the boundary of our sample, this is by no means necessary. Fronts can be nucleated in the interior of the network at a site of stress concentration or low network density. This prediction can also be tested since it is possible to visualize moving fronts as we show in our experiments.

Fibrin deformability has become a rapidly developing area not only in the study of the mechanics of filamentous networks, but also in various practical biomedical applications. First, studies of compression of fibrin networks provide a basis for understanding what happens to salutary hemostatic clots and pathological obstructive thrombi \textit{in vivo}, when they are subject to natural deformations. There is a number of conceivable patho-physiological conditions, in which fibrin networks can experience compressive loads: vasoconstriction, arterial hydrodynamic blood flow, clot contraction, pulsation of a vessel wall or aneurism, muscular pressure, cardiac beats for intracardiac thrombi, lung motion during breathing, etc.  Next, fibrin has been widely used as a sealant in surgery and its hemostatic efficiency strongly depends on the ability to withstand mechanical loads, including compression. Finally, fibrin has been well-known for various tissue engineering applications in combination with cells and with other materials to replace damaged or diseased organs and tissues \cite{Ahmed}. Mechanical and structural insights into fibrin deformability in combination with a theoretical framework of studying fibrin mechanics provide a foundation for developing new treatment modalities. In particular, modeling of fibrin compressibility using the theory of foam mechanics opens an avenue for physically treating fibrin clots and thrombi as well as for developing new fibrin-based biomaterials with predictable properties. 

\section{ACKNOWLEDGMENTS}
We would like to thank Professor Paul A. Janmey for fruitful discussion and providing access to the rheometer and confocal microscope and Anne van Oosten for valuable technical assistance. We also thank Jianxu Chen and Professor Danny Chen for helping with image analysis. We acknowledge partial support from the National Science Foundation through grants NSF CMMI 09535448 and NSF CMMI 1066787 (PKP and XL), National Institute of Health grant NIH R01GM095959 (OVK and MSA), NIH 1U01HL116330 (OVK, RIL, JWW, and MSA), NIH HL090774 (JWW) and Walther Cancer Foundation (OVK).

\begin{appendix}
\section*{Appendix}
\subsection*{Structure of the Compression Front} \label{sec:tanhprof}
\begin{figure}[h]
\centering
\includegraphics[width=3in]{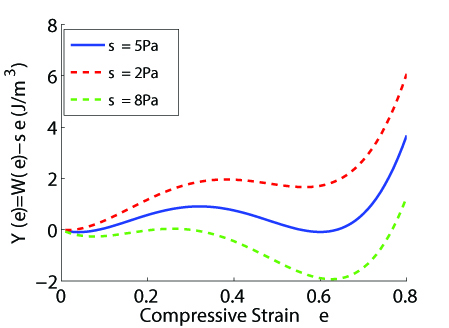}
\caption{Multi-well Gibbs free energy landscape for various stresses. At the plateau stress $\sigma = 5$Pa both wells are at equal height, meaning that the
straight and densified phases of the network can co-exist. For lower $\sigma$ the straight phase has lower energy and for higher $\sigma$ the densified
phase has lower energy.}
\label{energywell}
\end{figure}

In our analysis above we assumed that the compression front is sharp. However, our experiments reveal that it has a thickness over which there is a 
large gradient in the strain profile. Fortunately, these descriptions are connected and have been discussed at length in the literature~\cite{BookPhase}. 
Recall that we fitted the strain profile in our compressed sample using Eqn.  (\ref{eq:frontfit}).
In the following, we briefly describe why we use the strain profile given by Eqn. (\ref{eq:frontfit}). We follow the analysis in \cite{BookPhase}.  
Our network can exist in multiple phases at the same stress. Such materials have an up-down-up type stress-strain curve that can be described using 
\begin{equation}
 \sigma - \sigma_{0} = -\alpha(\varepsilon - \varepsilon_{*}) + \beta(\varepsilon - \varepsilon_{*})^{3},
\end{equation}
where $\sigma$ is the stress, $\varepsilon$ is the strain, and $\alpha$, $\beta$, $\sigma_{0}$ and $\epsilon_{*}$ are material parameters. {{If
we remember that $\sigma = \frac{\partial W}{\partial \varepsilon}$, where $W(\varepsilon)$ is a stored energy function, then it is not difficult to see that 
$W$ is a quartic function of $\varepsilon$ with two wells. The well at low strains corresponds to a network in which the fibers are straight, while the 
well at high strains corresponds to a densified network with buckled fibers.}} Without loss of generality we will now refer to $\sigma - \sigma_{0}$ as $\sigma$ and to $\varepsilon - \varepsilon_{*}$ as $\varepsilon$. As an example, 
$\sigma_{0} = 5$ Pa, $\epsilon_{*} = 0.32$, $\alpha = 51$ Pa and $\beta = 651$ Pa captures the compression curve for $E_{s}= 5$ MPa in Figure~\ref{phac}.
{{In Figure \ref{energywell} we have plotted the Gibbs free energy density ($\Psi = W - \sigma\varepsilon$) landscape using the above parameters for three different values of $\sigma$ to show which phase has lower energy at each stress.}} Since the motion of the compression front causes energy dissipation we augment this constitutive equation with a linear viscosity term and since interfaces 
cost energy we add a linear strain gradient term as follows~\cite{BookPhase}:
\begin{equation}
 \sigma = \alpha\varepsilon - \beta\varepsilon^{3} + \rho\nu\frac{\partial\varepsilon}{\partial t} 
 - \rho\lambda\frac{\partial^{2}\varepsilon}{\partial Z^{2}},
\end{equation}
where $\rho$ is the density of the material, $\nu$ is a viscosity and $\lambda$ accounts for the energy required to create an interface. We assume that
the compression front travels at velocity $v$ and define a new variable $\xi = Z - vt$. Since the stress $\sigma$ in our sample is constant we wish to
find a profile for $\varepsilon(z,t)$ that satisfies
\begin{equation}
 \rho\lambda\frac{d^{2}\varepsilon}{d\xi^{2}} + \rho\nu v\frac{d\varepsilon}{d\xi} + \alpha\varepsilon - \beta\varepsilon^{3} + \sigma = 0,
\end{equation}
with remote conditions $\varepsilon \rightarrow \varepsilon^{+}$ as $\xi \rightarrow \infty$ and $\varepsilon \rightarrow \varepsilon^{-}$ as 
$\xi \rightarrow -\infty$. It is easy to show that
\begin{equation}
 \varepsilon(z,t) = a + b\tanh(\frac{Z - vt - Z_{0}}{c}),
\end{equation}
gives such a profile for a particular choice of $a$, $b$, $c$ and $v$ that depend nonlinearly on $\alpha$, $\beta$, $\nu$, $\lambda$, $\rho$ and $\sigma$.
In particular, $a$ is a solution to the cubic $\sigma + 2\alpha a - 8\beta a^{3} = 0$, $b  = \sqrt{\frac{\alpha}{\beta} - 3a^{2}}$, 
$c = \frac{2\rho\lambda}{\beta b^{2}}$ and $v = \sqrt{\frac{3\beta abc}{\rho\nu}}$. We recover Eqn. (\ref{eq:frontfit}) by setting $v = 0$, justifying its use 
to fit the strain profile in our compressed samples. Note that in accordance with our assumption in Eqn. (\ref{eq3}) the front velocity $v$ in this augmented
theory depends on $\sigma$. A sharp front corresponds to the limit as $c \rightarrow 0$. In this limit the dissipation caused by the motion of the front can 
be captured by a kinetic equation like Eqn. (\ref{eq3}) as in \cite{BookPhase}.  

\end{appendix}


\section*{Addendum}
After the publication of our article \cite{Kim2015} online it was brought to our attention that there is some overlap in its content with an article 
published by Lakes {\it et al.}\cite{Lakes1993}. Unfortunately, we were unaware of this paper and did not cite it in our article. 
After carefully reading their paper, we provide a summary of the points of similarity and differences between the two papers below.
\begin{enumerate}
\item Both papers study, by a combination of experiment and theory, the compression behavior of fibrous materials. They both show that the measured 
stress-strain curve has three regimes -- a linear regime, plateau regime, and densified regime -- typical of foams \cite{Gibson}. However, the Scott 
industrial foam studied in \cite{Lakes1993} is a macroscopic reticulated open cell synthetic foam with a definite pore size, while the fibrin studied 
in \cite{Kim2015} is a microscopic isotropic network of natural fibers with a distribution of lengths and pore sizes that has not been examined in the 
context of theories for foams and cellular solids. 
\item Both papers show experimentally that in the plateau regime regions of densified material (with inter-fiber contact) co-exist with rarefied material,
and the fraction of densified material increases with increasing strain in the plateau regime. In \cite{Lakes1993} the bands of localized compression 
observed in experiments were viewed as regions of discontinuous strains alternating between two distinct values $\lambda_{+}$ and $\lambda_{-}$.
In \cite{Kim2015} we visualized a moving ``compression front" with densified material on top and rarefied material in the bottom. In addition, we showed 
that this front starts at the top of the sample and reaches the bottom at compressive strains greater than 0.6. 
\item Both papers argue that the plateau in the stress-strain curve can be theoretically explained by accounting for the co-existence of the rarefied and
densified phases of material at the same stress $\sigma$, but distinct values of the stretch $\lambda$ in each phase. Lakes {\it et al.}\cite{Lakes1993}
argued that their strain discontinuities are analogous to phase boundaries in continuum theories of phase transitions, and that minimization of total 
potential energy in the presence of these phase boudaries leads to the stress plateau. Without prior knowledge of the work 
in \cite{Lakes1993} we explained the plateau in the stress-strain curve of compressed fibrin by the quasistatic motion of the ``compression front", 
or phase boundary, visualized in our experiments.     
\item Both papers point out that some of the energy dissipation observed in experiments can be partly attributed to the dissipative motion of phase 
boundaries. While this was a qualitative statement in \cite{Lakes1993} we have quantitatively deduced a mobility parameter for phase 
boundaries in compressed fibrin networks by fitting our measurements of the storage and loss moduli at various compressions with analytical expressions 
derived using a continuum theory of phase transitions \cite{Abeyaratne}.    
\item Both papers point out that the phase boundary is not a sharp interface but has a thickness over which there is a large gradient in the strain \cite{Kim2015}. In \cite{Lakes1993} it was observed that the width of the bands depended on cell size. The authors mentioned that the length scale of the bands  
could be accounted for via models which incorporate strain gradients. In \cite{Kim2015} we quantitatively fit the strain profile observed in our 
experiments with analytical expressions derived from an augmented theory of phase transitions incorporating strain gradients \cite{Abeyaratne}. We show that
the thickness of the phase boundary is many times larger than the average length of a fiber between branch points.  
\end{enumerate}
The results of \cite{Lakes1993} are important and relevant to the experimental and theoretical study of foams and our paper, and we duly cite the article 
through this addendum.

\end{document}